\documentclass[a4paper,11pt]{article}
\pdfoutput=1

\usepackage{jheppub} 

\usepackage[T1]{fontenc} 

\usepackage{amssymb}
\usepackage[normalem]{ulem} % delete me
\usepackage{ulem}

\def\WWW{e^+\nu_e\, \mu^-\bar{\nu}_\mu\, \tau^+\nu_\tau\, b\bar{b}}

\def\tb{\bar{t}}

\def\GeV{\,\textrm{GeV}}

\title{NLO QCD predictions for off-shell $\boldsymbol{t\bar{t}W}$
  production in association with a light jet at the LHC}

\author[a, \,b]{Huan-Yu Bi,}
\author[c]{Manfred Kraus,}
\author[d]{Minos Reinartz\,}
\author[d]{and Malgorzata Worek\,}

 \affiliation[a]{Center for High Energy Physics, Peking University, Beijing 100871, China}
 \affiliation[b]{School of Physics, Peking University, Beijing 100871, China}
 \affiliation[c]{Departamento de F\'{i}sica Te\'{o}rica, Instituto de F\'{i}sica, \\ Universidad Nacional Aut\'{o}noma de M\'{e}xico, Cd. de M\'{e}xico C.P. 04510, M\'{e}xico}
\affiliation[d]{Institute for Theoretical Particle Physics
and Cosmology, RWTH Aachen University, \\D-52056 Aachen, Germany}

\emailAdd{bihy@pku.edu.cn}
\emailAdd{mkraus@fisica.unam.mx}
\emailAdd{minos.reinartz@rwth-aachen.de}
\emailAdd{worek@physik.rwth-aachen.de}

\abstract{In view of the persisting tension between theoretical predictions and the LHC data for the $pp \to t\bar{t}W^\pm$ production process, we present the state-of-the-art full off-shell NLO QCD result for  $pp \to t\bar{t}W^+\, j+X$.  We concentrate on the multi-lepton decay channel at the LHC with $\sqrt{s}=  13$ TeV.  In our calculation off-shell top quarks and gauge bosons are described by Breit-Wigner propagators,  furthermore, double-, single- as well as non-resonant top-quark contributions along with all interference effects are consistently incorporated at the matrix element level.  We present results for both integrated and differential fiducial cross sections for various renormalisation and factorisation scale settings and different PDF sets.  With a fairly inclusive choice of cuts and regardless of the scale and PDF choice, non-flat differential ${\cal K}$-factors  are obtained for many observables that we have examined. Since from an experimental point of view, both processes  $pp \to t\bar{t}W^\pm j+X$  and $pp\to t\bar{t}W^\pm +X$  consist of similar final states we investigate the effect of additional jet activity on the integrated and differential fiducial cross sections. For this purpose, the normalised differential distributions for $pp \to \WWW \,j+X$ and $pp \to \WWW +X$ are compared. The theoretical results for the latter process are  also recalculated.}

\dedicated{\rm TTK-23-11, P3H-23-029}


\begin{document} 

\maketitle
\flushbottom

% =============================================
%
\section{Introduction}
\label{sec:introduction}
%
% =============================================

An associated production of a top-quark pair and a $W$ gauge boson, denoted
as $pp \to t\bar{t}W^\pm$, is a rare process at the Large Hadron Collider (LHC). With a small  
cross section, compared to other associated productions such as $pp\to 
t\bar{t}j$, $pp\to t\bar{t}jj$ or $pp\to t\bar{t}b\bar{b}$, it is  
nevertheless an interesting process for several reasons. First, it
can  help to measure various properties of the top quark and the $W$ boson 
with  higher precision. This helps in the further verification of the consistency of the Standard Model (SM) as well as in the search for new physics effects that may manifest themselves in small corrections to the SM predictions.  Furthermore, the charge asymmetry of the top quark $(A_c^t)$ and its decay products $(A_c^\ell, A_c^b)$ in the $t\bar{t}W^\pm$ production process plays an important role at the LHC as these quantities are predicted to be larger than in pure $t\bar{t}$ production \cite{Maltoni:2014zpa,Bevilacqua:2020srb}. An enhancement  in the measured values of  $A_c^t$,  $A_c^\ell$ and $ A_c^b$ could expose possible new physics contributions like for example the presence of a massive color octet vector boson  (axigluon) that in general can couple in a different way to light and heavy quarks, see e.g. Refs. \cite{Ferrario:2009bz,Frampton:2009rk,Maltoni:2014zpa}.  Second, $pp \to  t\bar{t}W^\pm$ contributes to the production of  same-sign leptons. Both ATLAS and CMS experiments  have  reported sustained excesses
in the production of  same-sign leptons in the corners of phase-space where the production of top-quark pairs in association with a $W$  boson is one of  the dominant processes \cite{ATLAS:2016dlg,CMS:2016mku,CMS:2017tec,ATLAS:2017tmw,CMS:2017ixv,ATLAS:2019fag,CMS:2020cpy}. Thus, understanding as precisely as possible rare processes  involving top
quarks is of  high  interest at the LHC. In fact, $pp \to t\bar{t}W^\pm$
is one of the most important backgrounds in the search for new physics
where same-sign leptons are the main signature, such as the
production of  supersymmetric particles \cite{Barnett:1993ea,Guchait:1994zk,Baer:1995va,Maalampi:2002vx,Dreiner:2006sv} or vector-like quarks \cite{delAguila:1998tp,delAguila:2000aa,delAguila:2000rc,Aguilar-Saavedra:2013wba}. Such a  signature is also present, among others, in models with universal extra dimensions \cite{Cheng:2002ab},
heavy  top-quark partners \cite{Contino:2008hi,DeSimone:2012fs} and an extended Higgs-boson sector \cite{vonBuddenbrock:2016rmr,vonBuddenbrock:2017gvy,vonBuddenbrock:2018xar,Buddenbrock:2019tua}. In addition, same-sign leptons  are considered a key feature in searches for heavy  Majorana neutrinos \cite{Almeida:1997em} as well as for $tt$ and $\bar{t}\bar{t}$ resonances \cite{Bai:2008sk,Berger:2011ua}.  Third, the $pp \to  t\bar{t}W^\pm$ process plays a vital role in many
measurements of SM  processes  such as $pp\to t\bar{t}$  production in  
association with a Higgs boson \cite{ATLAS:2018mme,CMS:2018uxb,ATLAS:2023cbt,CMS:2020cga} or the  production of four top quarks,  $pp\to  t\bar{t}t\bar{t}$, which  has been  recently observed by both ATLAS and CMS \cite{ATLAS:2023ajo,CMS:2023ica}.   In  the recent measurements of $pp \to 
t\bar{t}H$ and $pp\to t\bar{t} W^\pm$ production in multi-lepton final 
states  tension has been seen  in the modelling of 
the final-state kinematics in the  phase-space regions dominated by 
$t\bar{t} W^\pm$  production \cite{ATLAS:2019nvo}. 

Both ATLAS and CMS  have measured  $pp \to t\bar{t}W^\pm$ with $8$ TeV and $13$ TeV  centre-of-mass energy data \cite{CMS:2015uvn,CMS:2017ugv,CMS:2022tkv,ATLAS:2015qtq,ATLAS:2016wgc,ATLAS:2019fwo}. 
Multipurpose Monte Carlo event generators and various theoretical predictions available in the literature for this process,  which are  employed by both collaborations, are not  able to very precisely describe  the current $pp \to t\bar{t}W^\pm$ data 
\cite{CMS:2022tkv,ATLAS-CONF-2023-019}. In both cases the measured inclusive and integrated fiducial cross sections are consistently larger than SM predictions. In the case of ATLAS in addition various absolute and normalised  differential cross section distributions have been measured. Furthermore, the results in the $3\ell$ channel have been compared to the state-of-the-art fixed-order NLO QCD predictions for $pp \to \ell^+ \nu_{\ell} \, \ell^- \bar{\nu}_\ell \, b\bar{b} \, \ell^\pm \nu_\ell$ with full off-shell effects included \cite{Bevilacqua:2020pzy}.  This prediction has been calculated at parton-level and corrected to particle-level through bin-by-bin scaling factors that account for non-perturbative effects, such as multi-parton interactions and hadronisation. All absolute differential cross section measurements have shown a rather significant problem in the normalisation, that is expected from the result of the inclusive cross section measurement. For normalised differential cross-section distributions, the agreement with the data is better, as evidenced by $\chi^2$ $p$-values above $0.5$ for almost all observables.
Nevertheless, some observables still have shown tensions in various phase-space regions. In addition,  NLO plus parton shower predictions with additional hard jets as described by the matrix element and obtained by  merging of  $pp \to t\bar{t}W^\pm$  and $pp \to t\bar{t}W^\pm j$ processes \cite{Frederix:2021agh} are in better agreement with data than exclusive NLO plus parton shower predictions for the $pp \to t\bar{t}W^\pm$ process alone. This demonstrates the importance of additional light jet emissions in correctly modeling the $pp\to t\bar{t}W^\pm$  process. In all these measurements, the accumulated amount of data has been the largest source of uncertainty across most bins of the measurement. The most important systematic uncertainties, on the other hand, have been identified as arising from modelling of the $pp\to t\bar{t}W^\pm$ process. Furthermore,
a search for the leptonic charge asymmetry in $t\bar{t}W^\pm$ production in final states with three leptons at the LHC with $13$  TeV using an 
integrated luminosity of $139 ~{\rm fb}^{-1}$ has been presented by the ATLAS collaboration  \cite{ATLAS:2023xay}. Specifically,  the $A^\ell_c$ charge asymmetry was extracted together with the 
normalisations for the $pp\to t\bar{t}W^\pm$ and $pp \to t\bar{t}Z$ background processes. At reconstruction level, the asymmetry was found to be $A^\ell_c= -0.123 \pm 0.136 ~{\rm (stat.)}  \pm 0.051 ~{\rm (syst.)}$. An unfolding procedure has been applied to convert the result at the reconstruction level into the charge-asymmetry value in a fiducial volume at particle level with the result of $A^\ell_c = - 0.112 \pm 0.170 ~{\rm (stat.)} \pm 0.054 ~{\rm (syst.)}$. Due to large uncertainties both results are consistent with the SM expectations. The most relevant systematic uncertainties affecting this search are again coming from the modelling uncertainties of the $t\bar{t}W^\pm$ and $t\bar{t}Z$ MC processes in the $3\ell$ channel. Furthermore, both the reconstruction- and particle-level results are severely limited by the statistical uncertainties of the data. The collected data of the LHC Runs 1 and 2 and the  increased integrated luminosity that is  expected in the LHC Run 3, and finally the high luminosity phase of the LHC,  will deliver sufficient amount of data to probe $pp \to t\bar{t}W^\pm$ production  and top-quark  properties in this process in greater detail. Thus, the quest to better understand the dynamics of the $t\bar{t}W^\pm$ process is still ongoing.

The experimental signature of the  $pp \to t\bar{t}W^\pm$ process, 
characterized by the presence of charged  leptons  (electrons or muons), 
$b$-tagged jets, light jets as well as missing  transverse momentum, is 
very challenging. Large or in many cases not so well known 
backgrounds,  particularly from $pp \to t\bar{t}$ and  $pp \to t\bar{t}H$
as well as  from single  top-quark  processes, e.g. $pp \to tW(b)$ and $pp \to tWH(b)$  need to be precisely estimated to be subtracted from the small
signal process. In  addition, proper modelling of the kinematic properties of
the final-state particles is vital. The ability to correctly describe the
single top-quark contribution is crucial and might prove to be a
bottleneck, given the current state of predictions used at the LHC for the
$pp \to tW(b)$ process \cite{ATLAS:2018ivx,ATLAS:2021pyq}. To top it all off,  the production of a $t\bar{t}$ 
pair in  association with a $W^\pm$ boson is quite a peculiar  process. At
LO in the perturbative expansion this process can only  occur
via a  $q\bar{q}^{\,\prime}$ annihilation. At NLO, 
the $qg$   channels opens up, yet the gluon-gluon fusion production is  not
reachable until NNLO is included. This 
is in contrast to other $t\bar{t}$ associated  boson productions like for 
example $pp \to t\bar{t}Z$, $pp \to t\bar{t}H$  or $pp \to t\bar{t}\gamma$,
where  bosons can also couple to the top-quark pair in  the $gg \to
t\bar{t}Z/t\bar{t}H/t\bar{t}\gamma$  subprocess at LO. In the absence of NNLO QCD predictions for $pp\to 
t\bar{t}W^\pm$  production, the only other possibility to include the $gg$
subprocess is to study the $pp\to  t\bar{t}W^\pm j$ process at NLO in QCD or provide 
the multi-jet merged samples, which should capture parts of NNLO QCD 
contributions since  hard  non-logarithmically enhanced radiation is 
properly described by corresponding matrix elements. 

The goal of this paper is to provide the state-of-the-art fixed-order NLO QCD prediction for the $pp \to \WWW j$ process, for the sake of brevity referred to as $pp \to t\bar{t}W^+ j$, for the LHC Run 2 center-of-mass system energy of $\sqrt{s}=13$ TeV. In our computations off-shell top quarks and $W$ gauge bosons are described by Breit-Wigner propagators, furthermore, double-, single- as well as non-resonant contributions along with all interference effects are consistently incorporated at the matrix element level. We scrutinise the size of higher-order corrections and theoretical uncertainties in such a complex environment. We additionally address the choice of a judicious renormalisation and factorisation scale setting and the size of PDF uncertainties. Finally, a comparison of LO and NLO fiducial cross sections for the $t\bar{t}W^+$ and $pp\to t\bar{t}W^+ j$ processes is performed. 

As a final comment, we note that for the inclusive $pp \to t\bar{t}W^\pm$ process, with stable top quarks and $W$ gauge boson, NLO QCD corrections have already been around for some time \cite{Hirschi:2011pa,Maltoni:2014zpa,Maltoni:2015ena}. Theoretical predictions for  $pp \to t\bar{t}W^\pm$ at NLO in QCD have been also  matched to parton shower Monte Carlo programs using either the \textsc{Powheg} method or the \textsc{Mc@Nlo} framework \cite{Hirschi:2011pa,Garzelli:2012bn,Maltoni:2014zpa,Maltoni:2015ena,FebresCordero:2021kcc}.  In all cases top-quark and $W$ gauge boson decays have been treated in
the parton shower approximation omitting NLO $t\bar{t}$ spin correlations. In Ref. \cite{Campbell:2012dh} improved fixed-order predictions for  $pp \to t\bar{t}W^\pm$ have been presented, where NLO QCD corrections to the production and decays of top quarks and $W$ gauge bosons have been included with full spin correlations utilising the narrow-width approximation. A further step towards a more precise modelling of the on-shell $pp \to t\bar{t}W^\pm$  process has been achieved by including either NLO electroweak corrections \cite{Frixione:2015zaa} and  subleading electroweak corrections \cite{Dror:2015nkp,Frederix:2017wme,Frederix:2020jzp} or by incorporating soft gluon resummation effects with next-to-next-to-leading logarithmic (NNLL)  accuracy \cite{Li:2014ula,Broggio:2016zgg,Kulesza:2018tqz,Broggio:2019ewu,Kulesza:2020nfh}. They  are evaluated by adding to the NLO result the ${\cal O} (\alpha_s^2)$ term 
of the expansion of the NNLL soft gluon resummation of the cross section. The soft gluon resummation concerns, however, only the Born ($q\bar{q}^{\, \prime} \to t\bar{t}W^\pm$) process. Consequently, such predictions do not substantially improve NLO results for the $pp \to t\bar{t}W^\pm$  process. In addition, multi-jet  merged predictions are available
in the literature for $pp\to t\bar{t}  W^\pm$ albeit only for  on-shell top 
quarks and $W^\pm$ gauge bosons \cite{vonBuddenbrock:2020ter,Frederix:2021agh}. In Ref.  \cite{Frederix:2021agh} LO top-quark and $W$ gauge boson decays have been incorporated within the NWA. Consecutively these predictions have been matched to 
parton  shower programs. Single  top-quark contributions, however,  are still not 
included  and $t\bar{t}$ spin  correlations are  incorporated
with LO accuracy only. Recently, NLO QCD corrections to $t\bar{t}W^\pm$ 
with full off-shell effects included   appeared in the literature albeit for the multi-lepton final state only \cite{Bevilacqua:2020pzy}. They have been compared to  NWA results  to assess the size of  nonfactorisable NLO QCD corrections for the integrated fiducial cross section and various differential distributions \cite{Bevilacqua:2020pzy}. Furthermore, their impact on the ${\cal R}= t\bar{t}W^+/ t\bar{t}W^-$ cross section ratio and $A^t_c, \, A^\ell_c,\, A^b_c$ has been studied \cite{Bevilacqua:2020srb}.  Independent computations \cite{Denner:2020hgg} not only confirmed the results presented in Ref. \cite{Bevilacqua:2020pzy} but also performed a comparison between the full result and the one obtained with the help of the double-pole approximation. In addition, NLO QCD and electroweak corrections to the full off-shell $pp\to t\bar{t}W^\pm$ production  have been combined for the three-charged-lepton channel \cite{Denner:2021hqi}. Finally, in Ref. \cite{Bevilacqua:2021tzp} full off-shell and NWA theoretical predictions for $t\bar{t}W^\pm$  have been compared to parton shower matched results from Ref. \cite{FebresCordero:2021kcc}. In detail, not only the dominant NLO QCD  contributions at the perturbative orders ${\cal O}(\alpha_s^3 \alpha^6)$ and ${\cal O}(\alpha_s \alpha^8)$ have been studied in Ref. \cite{Bevilacqua:2021tzp}, but also they have been  combined to approximately incorporate the full off-shell effects in the NLO  computation of on-shell $pp \to t\bar{t}W^\pm$ matched to parton showers.

The content of the present work is organized as follows. In Section \ref{sec:comptsetup} we briefly describe  our computational framework. In Section  \ref{sec:inputparameters} we provide the theoretical setup for LO and NLO QCD predictions. Results for the $pp \to t\tb W^+ j + X$ process in the multi-lepton decay channel are presented in Section \ref{sec:fiducialxs} and Section \ref{sec:differential}. In Section \ref{sec:comptottw} we perform a comparison between the $pp \to t\tb W^+ j + X$ and  $pp \to t\tb W^+ + X$ processes. Finally, our findings are summarised in Section \ref{summary}.

% =============================================
%
\section{Details of the calculation}
\label{sec:comptsetup}
%
% =============================================
%
%=============================================
\begin{figure}[t]
  \begin{center}
    \includegraphics[width=\textwidth, trim = 25 26.1cm 25 0cm, clip]{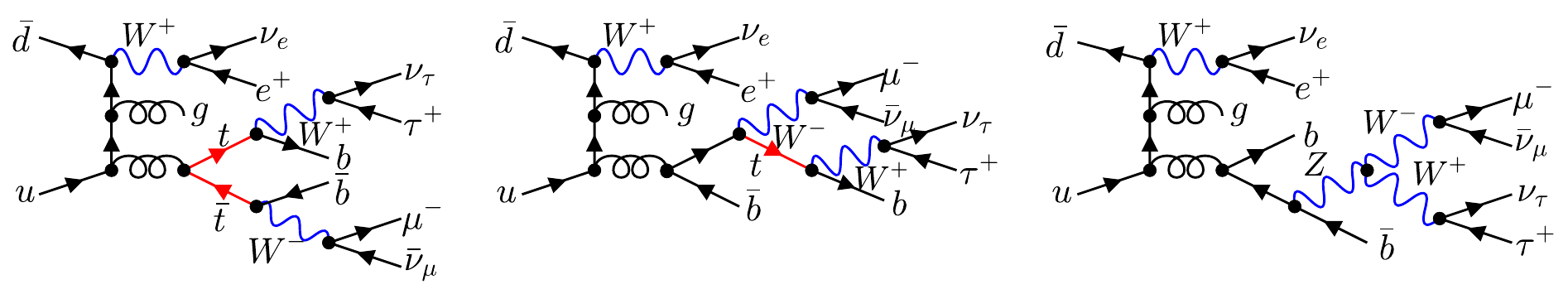}
    \includegraphics[width=\textwidth, trim = 25 26.2cm 25 0cm, clip]{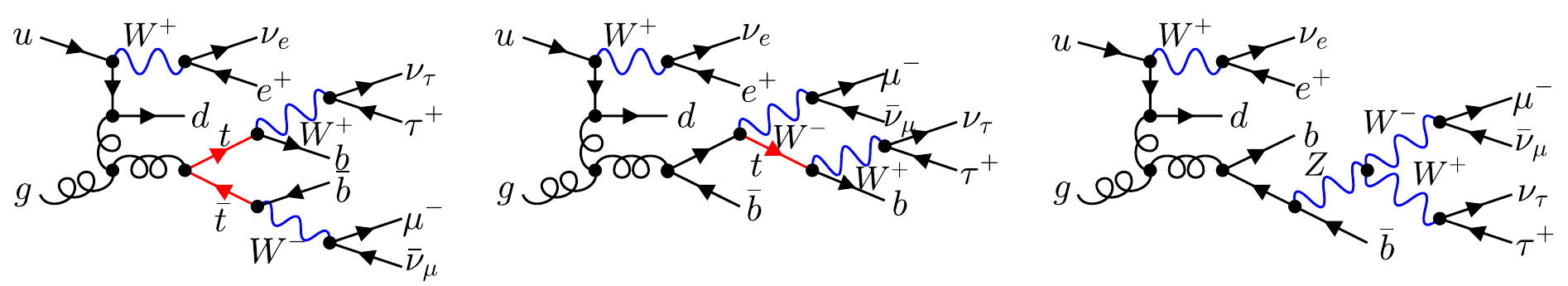}
\end{center}
\caption{\label{fig:feynmandiagrams} \it 
  Representative Feynman diagrams with double- (left), single- (middle) and no top-quark resonances (right) contributing to the $ u\bar{d} \to \WWW\, g $ and $gu \to \WWW \, d$ partonic subprocess at leading order.  We include all three types of contributions in our calculations together with their interference effects. The \textsc{FeynGame} program \cite{Harlander:2020cyh} is employed to draw Feynman diagrams.}
 \end{figure}
%=============================================
%

In $pp$ collisions at the LHC the $\WWW j$  final state is produced via the scattering of a quark and gluon, $qg \to \WWW q^{\, \prime}$, or a $q\bar{q}^{\, \prime}$ pair, $q\bar{q}^{\, \prime} \to \WWW g$, where $q=u,d,c,s$. The contributions to the tree-level squared amplitude at  ${\cal O} (\alpha_s^3 \alpha^6)$  comprise  Feynman  diagrams with two top-quark resonances, one top-quark resonance and no top-quark resonances as well as all their interference terms. Representative LO Feynman diagrams for the $ u\bar{d} \to \WWW\, g $ and $gu \to \WWW \, d$ partonic subprocess  are presented in Figure \ref{fig:feynmandiagrams}. At LO, we distinguish 12 partonic subprocesses, each of which has 1868 Feynman diagrams. We emphasis here that we do not use Feynman diagrams in our calculations. Instead, matrix elements and integrated fiducial cross sections are obtained with the Dyson-Schwinger recursive algorithm \cite{Draggiotis:1998gr,Kanaki:2000ey,Draggiotis:2002hm}. Nevertheless, we present the number of Feynman diagrams as a measure of the complexity.
Even though we treat $b$ quarks as massless partons there are Higgs-boson-exchange Feynman diagrams. Using the LO calculation we have checked, however, that such contributions are below permill level. Therefore, they are  within our Monte Carlo integration errors and can be safely omitted throughout the calculation. For the purpose of this test we set $m_H=125\GeV$ and $\Gamma_H=4.07\cdot10^{-3}\GeV$.  To regularise intermediate top-quark resonances in a gauge-invariant way we employ the complex-mass scheme, see e.g. Refs.  \cite{Denner:1999gp,
  Denner:2005fg, Bevilacqua:2010qb,Denner:2012yc,Denner:2014zga}, which consistently describes off-shell top-quark contributions by the Breit-Wigner distribution. All matrix elements are evaluated using the complex top-quark mass, $\mu_t^2 = m_t^2 - im_t\Gamma_t$. In addition, the $W$ and $Z$ gauge bosons  are treated within the complex-mass scheme. The calculation of the scattering amplitudes is done within the \textsc{Helac-NLO} framework \cite{Bevilacqua:2011xh}, which comprises \textsc{Helac-Dipoles} and \textsc{Helac-1Loop} \cite{vanHameren:2009dr},  
  and with the help of the \textsc{Helac-Phegas} Monte Carlo program \cite{Cafarella:2007pc}. The latter is used to cross-check all LO results. The phase space integration is performed and optimised with  \textsc{Parni} and \textsc{Kaleu} \cite{vanHameren:2007pt,vanHameren:2010gg}.
%
% =============================================
\begin{table}[t!]
\begin{center}
\begin{tabular}{ccc}
\hline \hline  \\ [-0.4cm]
  \textsc{One-loop Correction Type} &
  & \textsc{Number of Feynman Diagrams}\\[0.2cm]
  \hline \hline  \\ [-0.4cm]
 {Self-energy}       & \quad \quad\quad\quad \quad 
 & 32308  \\ [0.2cm]
 {Vertex}               &&  24854  \\ [0.2cm]
 {Box-type}           & & 14218   \\ [0.2cm]
 {Pentagon-type}  & &  8020   \\ [0.2cm]
 {Hexagon-type}   && 3346  \\[0.2cm]
{Heptagon-type}   & & 812   \\ [0.2cm]
{Octagon-type}     & & 56  \\ [0.2cm]
  \hline \hline \\[-0.4cm]
{Total number}      & &  83614  \\[0.2cm]
   \hline \hline
 \end{tabular}
\end{center}
\caption{\label{table:one-loop} \it
The number of one-loop Feynman diagrams for the subprocess
$u \Bar{d} \to \WWW\, g$ at ${\cal O}(\alpha_s^4 \alpha^6)$ 
split by loop topology. 
The Higgs boson exchange contributions are neglected and the
Cabibbo-Kobayashi-Maskawa mixing matrix is kept diagonal.}
\end{table}
% =============================================

The virtual corrections are obtained from the interference of the sum of all one-loop diagrams with the Born amplitude. One can classify them into
self-energy, vertex, box-type, pentagon-type, hexagon-type, heptagon-type and octagon-type corrections. In Table \ref{table:one-loop} we provide the number of one-loop Feynman diagrams, that corresponds to each topology for the following partonic subprocess $u \Bar{d} \to \WWW \,  g$.  These numbers have been generated using the  \textsc{Qgraf} program \cite{Nogueira:1991ex}  and are provided to give a general sense of the complexity of our calculations.  The virtual corrections are calculated in the 't Hooft-Veltman \cite{THOOFT1972189} version of the dimensional regularization scheme. The singularities stemming from infrared divergent contributions are canceled by the respective poles in the integrated counter terms of the dipole subtraction approach. The finite contributions of the loop diagrams are evaluated numerically in $d = 4$ 
dimension. Throughout our calculations we monitor the numerical stability by checking Ward identities at every phase space point. The events which violate gauge invariance are recalculated with higher precision. The 
one-loop calculation is performed using
\textsc{Helac-1Loop}, which reweights a
leading-order event sample to the one-loop result. One loop integrals
are calculated using the \textsc{OPP} reduction method as implemented
in \textsc{CutTools} \cite{Ossola:2006us, Ossola:2007ax} in
combination with \textsc{OneLOop} \cite{vanHameren:2010cp} for the
calculation of one-loop scalar functions. We have cross-checked our results with the publicly available general purpose Monte Carlo program \textsc{MadGraph5-aMC@NLO} \cite{Alwall:2014hca,Hirschi:2011pa}. In practice,  we have checked the finite
parts along with the coefficients of the poles in $\epsilon$, i.e. $1/\epsilon^2$ and  $1/\epsilon$ of the virtual amplitudes  for a few phase-space points for the $u\bar{d} \to \WWW \, g$ and $gu \to \WWW \, d$  partonic subprocesses. 
%
%================================================
\begingroup
\renewcommand{\arraystretch}{1.5}
\begin{table}[t!]
\centering
\begin{tabular}{lccc}
    \hline\hline
     \textsc{Partonic} &  \textsc{Number Of} & \textsc{Number Of} & \textsc{Number Of}\\
      \textsc{Subprocess}&  \textsc{Feynman Diagrams} 
      & \textsc{CS Dipoles} & \textsc{NS Subtractions}\\[0.1cm]
     \hline\hline
     $gg\to\WWW \,qq^\prime$ & 
     16662 & 36&  9\\[0.1cm]
     $qq^\prime\to\WWW \, gg$ & 16662 & 40
     & 10\\[0.1cm]
     $qg\to\WWW q^\prime g$  &16662 &  36
     &  9\\[0.1cm]
     \hline\hline
     $qq^\prime\to\WWW \,qq^\prime$ & 6240 
     & 12
     &  3\\[0.1cm]
     $q\bar{q}\to\WWW \,qq^\prime$ & 6240
     &  8
     & 2\\[0.1cm]
     $qq^\prime\to\WWW \,b\bar{b}$ & 6240 
     & 16 &  4\\[0.1cm]
     \hline\hline
     $qq^\prime\to\WWW \,QQ^\prime$ &3120&
     8 &  2\\[0.1cm]
     $q\bar{q}\to\WWW \,QQ^\prime$ & 3120 
     &  4
     &  1 \\[0.1cm]
     $qQ^\prime\to\WWW \,qQ^\prime$ & 3120
     & 8 &  2 \\[0.1cm]
     \hline\hline
     $b\bar{b} \to\WWW \,qq^\prime$
     & 6240 & 12
     & 3 \\[0.1cm]
     $bq \to\WWW \,b Q^\prime$
     & 6240 &  16
     &  4 \\[0.1cm]
     $bq \to\WWW \,b q^\prime$
     & 6240 &  16
     & 4 \\[0.1cm]
          \hline\hline
\end{tabular}
\caption{\it 
The list of partonic subprocesses contributing to the subtracted real emissions for the $pp \to \WWW j$ process  Also shown are the number of Feynman diagrams, as well as the number of Catani-Seymour and Nagy-Soper subtraction terms. We denote here $q,Q=u,d,c,s$ but  $q$ and $Q$  label  quarks from different generations.}
\label{table:fd}
\end{table}
\endgroup
%==============================================

The real emission corrections to the LO process arise from tree-level amplitudes with one additional parton, either an additional gluon or a $q\bar{q}$ pair replacing a gluon. All possible contributions of the order of ${\cal O} (\alpha_s^4\alpha^6)$ can be divided into $12$ categories which are given in Table \ref{table:fd}. Here  $q,Q= u,d,c,s$ but $q$ and $Q$ label quarks from different generations, i.e. $ u c \to \WWW \, d c$. We use an exclusive setup by requiring exactly two $b$-jets in the final state, where we define a $b$-jet using the charge-aware $b$-jet tagging as described in Ref. \cite{Bevilacqua:2021cit}.  Although the overall contribution of subprocesses with one or two bottom quarks in the initial state is numerically negligible (i.e. it is well below the final Monte Carlo error for the full NLO QCD result both at the integrated and differential cross-section level), we include it in our calculations for consistency reasons. This results in  $114$ subprocesses in total, instead of $94$ in the case without $b$ quarks in the initial state, that need to be included in the real emission contribution of the full NLO result. We employ the Catani-Seymour dipole subtraction scheme as defined in Ref. \cite{Catani:1996vz} to extract the soft and collinear infrared singularities and to combine them with the virtual corrections. In detail, we use the formulation presented in Ref. \cite{Catani:2002hc} for massive quarks. The latter has been further extended to arbitrary helicity eigenstates of the external partons in Ref. \cite{Czakon:2009ss} and is now implemented in \textsc{Helac-Dipoles}.  Furthermore, we employ the Nagy-Soper subtraction scheme \cite{Bevilacqua:2013iha}, which makes use of random polarisation and colour sampling of the external partons and it is also available in  \textsc{Helac-Dipoles}. Two independent subtraction schemes allow us to 
cross-check the correctness of the real corrections by comparing the two results. As an example we also display in Table \ref{table:fd} the total number of  the Catani-Seymour dipoles and the Nagy-Soper subtraction terms that can be constructed for the corresponding subprocess. In each case, four times less terms are needed in the Nagy-Soper subtraction scheme compared to the Catani-Seymour one. The difference corresponds to the total number of possible spectators that are only relevant in the Catani-Seymour case.   To further cross-check our results a restriction on the phase space of the subtraction term is used for both Catani-Seymour \cite{Bevilacqua:2009zn} and Nagy-Soper schemes \cite{Czakon:2015cla}.  Independence of said restriction is then checked to validate the computation. Moreover, we have ensured that the poles in $\epsilon$ occurring in the ${\cal I}$-operator and those appearing in the virtual part cancel  by explicitly checking a few phase-space points. By combining virtual and real corrections, singularities connected to collinear configurations in the final state as well as soft divergences in the initial and final states cancel out automatically for collinear-safe observables defined with the help of a jet algorithm. Singularities associated with collinear initial-state splittings are removed by factorization through  parton distribution function redefinitions. 

LO and NLO results for the $pp \to \WWW j$ process are stored in modified Les Houches files \cite{Alwall:2006yp}, which are then converted to \textsc{Root} Ntuple files \cite{Antcheva:2009zz}. We  follow the ideas outlined in  Ref. \cite{Bern:2013zja} and store  “events” with supplementary matrix element and PDF information. This allows us to obtain theoretical predictions  for different scale settings and PDF choices by simply reweighting previously created files. To this end we use the in-house software \textsc{HEPlot} \cite{Bevilacqua:HEPlot}.  Storing “events” has clear advantages when either different observables, observable's ranges, binning or more exclusive selection cuts are required for example when comparing theoretical predictions with LHC data or performing other  phenomenological studies. 

% =============================================
%
\section{Computational setup}
\label{sec:inputparameters}
%
% =============================================
%

We calculate NLO QCD corrections to the $pp \to \WWW j$ process 
for the LHC Run 2 energy of $\sqrt{s}=13$ GeV. Specifically, $\alpha_s$ corrections to the born-level process at ${\cal {O}}(\alpha_s^3 \alpha^6)$ are evaluated. For comparison we also consider  the $pp \to \WWW+X$ process at NLO in QCD, i.e. $\alpha_s$ corrections to the born-level process at perturbative order ${\cal {O}}(\alpha_s^2 \alpha^6)$. 
 We only simulate decays of the weak bosons to different lepton generations to avoid virtual photon singularities stemming from quasi-collinear $\gamma \to \ell^+\ell^-$ decays and to reduce the overall number of Feynman diagrams that we have to compute.  However, the  $Z/\gamma \to \ell^+\ell^-$ contribution is at the  per-mille level for our cuts, as checked by an explicit leading order calculation\footnote{In detail, we made comparisons with the following  LO processes $pp \to e^+\nu_e\, \mu^-\bar{\nu}_\mu\, e^+\nu_e\, b\bar{b} \,j$, $pp \to e^+\nu_e\, e^-\bar{\nu}_e\, e^+\nu_e\, b\bar{b} \, j$ and $pp \to e^+\nu_e\, e^-\bar{\nu}_e\, \mu^+\nu_\mu\, b\bar{b} \,j$.}. Since $e^\pm$ and $\mu^\pm$ are treated in the same way with respect to phase space cuts, the full cross section for the $pp\rightarrow l^+ \nu_l \,l^-\bar{\nu}_l \, l^+ \nu_l  \, b \bar{b}\, j + X$ process, where $l^{\pm}=e^{\pm},\mu^{\pm}$,  can be obtained by multiplying the results from this paper  with a lepton-factor of $4$. In line with our previous work for the simpler case of $pp \to e^+ \nu_e \, \mu^- \bar{\nu}_\mu \, e^+ \nu_e \, b\bar{b} +X $ \cite{Bevilacqua:2020pzy} also here we keep the  Cabibbo-Kobayashi-Maskawa  mixing matrix in the diagonal form. As recommended by the PDF4LHC working group we employ three sets of parton density functions (PDFs) for the use at the LHC \cite{PDF4LHCWorkingGroup:2022cjn}. In particular, our default PDF set is NNPDF3.1 \cite{NNPDF:2017mvq}. In addition we also provide LO and NLO results for CT18 \cite{Hou:2019efy} and MSHT20 \cite{Bailey:2020ooq}. As there is no LO CT18 PDF set, we use in this case CT14 \cite{Dulat:2015mca} instead. The corresponding prescription from each PDF fitting group is employed to provide the $68\%$ confidence level
PDF uncertainties. The running of  $\alpha_s$ with two-loop (one-loop) accuracy at NLO (LO) is provided by  the LHAPDF 6 library  \cite{Buckley:2014ana} including five active flavours. Furthermore,  we use the following SM parameters as provided for example in Ref. \cite{ParticleDataGroup:2020ssz}
\begin{align}
  &G_{ \mu}=1.166378 \cdot 10^{-5} ~\textrm{GeV}^{-2}\;,
  &m_t = 172.5~\textrm{GeV}\;, \\
  &m_W = 80.379~\textrm{GeV}\;,
  &\Gamma_W^{\textrm{NLO}} = 2.0972~\textrm{GeV}\;, \\
  &m_Z = 91.1876~\textrm{GeV}\;,
  &\Gamma_Z^{\textrm{NLO}} = 2.5074~\textrm{GeV}\;. 
\end{align}
All other leptons and quarks (including the $\tau$ lepton and the $b$ quark) are treated as massless. The top-quark width is calculated according to Ref. \cite{Denner:2012yc} and is treated as a fixed parameter throughout this work. Moreover, its value corresponds to a fixed scale setting of $\mu_0=m_t$. When computed for unstable $W$ bosons with $m_b=0$ GeV the LO and NLO top-quark width reads
\begin{align}
 &\Gamma^{\textrm{LO}}_{t, \,\textrm{off-shell}} = 1.45766~\textrm{GeV}\;, 
 &\Gamma^{\textrm{NLO}}_{t, \,\textrm{off-shell}} = 1.33254~\textrm{GeV}\;.
\end{align}
Finally,  the electromagnetic coupling $\alpha$ is calculated from the Fermi constant $G_\mu$ via
\begin{equation}
\alpha_{G_\mu}=\frac{\sqrt{2}}{\pi} \,G_\mu \,m_W^2  \,\sin^2\theta_W \,,
~~~~~~~~~~~{\rm
  where} ~~
\sin^2\theta = 1-\frac{m_W^2}{m_Z^2}\,,
\end{equation}
The fixed-order perturbative calculation in QCD introduces a dependence on the renormalisation scale, $\mu_R$, and the factorization scale, $\mu_F$. To
estimate the uncertainty introduced by the choice of these parameters
we vary the two scales around a central scale $\mu_0$ independently 
by a factor of 2 while limiting their ratio between 0.5 and 2.
%
%\begin{equation*}
%\frac{1}{2} \le \frac{\mu_R}{\mu_F} \le 2
%\end{equation*}
%
This results in the evaluation of the following seven scale variations
\begin{equation}
\label{scan}
\left(\frac{\mu_R}{\mu_0}\,,\frac{\mu_F}{\mu_0}\right) = \Big\{
\left(2,1\right),\left(0.5,1  
\right),\left(1,2\right), (1,1), (1,0.5), (2,2),(0.5,0.5)
\Big\} \,.
\end{equation}
For the central value of the scale, $\mu_0$, we consider three possibilities, with the default being set to
\begin{equation}
\mu_0 =\mu_R= \mu_F =  \frac{H_T}{2}\,, 
\end{equation}
where $H_T$ is the scalar sum of the transverse momenta of all decay products. It is defined according to 
\begin{equation}
H_T = p_T(e^+ )+ p_T(\tau^+)+ p_T(\mu^-) +p_{T}^{miss} + p_{T} (b_1) +
p_{T} (b_2) + p_T(j_1)\,,
\label{eq:ht}
\end{equation}
where $p_T^{miss}$ is the missing transverse momentum from escaping neutrinos given by
\begin{equation}
  p_T^{miss} = |\vec{p}_{T} (\nu_e) + \vec{p}_{T} (\bar{\nu}_\mu) + \vec{p}_{T} (\nu_\tau) |\,.  
\end{equation}
Moreover, $b_1$  and $b_2$ are the two $b$-jets whereas $j_1$ denotes the light jet (in the case of two resolved light jets that pass all cuts, the one with the larger $p_T$ is selected). In addition, we provide results for the following fixed scale setting 
\begin{equation}
\mu_0=\mu_R= \mu_F =  m_t+\frac{m_W}{2} \,,
\end{equation}
as well as for yet another dynamic scale choice 
\begin{equation}
\mu_0=\mu_R= \mu_F =  \frac{E_T}{2}\,.
\end{equation}
The latter is given by
\begin{equation}
E_T = \sqrt{m_t^2 + p_{T}^2(t)} + \sqrt{m_t^2 + p_{T}^2 (\tb\,)} +
\sqrt{m_W^2 + p_{T}^2 (W)} + p_T(j_1) \,.
\label{eq:et}
\end{equation}
Top quarks and $W$ gauge boson momenta are reconstructed from decay products by finding the decay history that minimizes the ${\cal Q}$ value 
\begin{equation}
{\cal Q} = \left| M(t) - m_t \right| + \left| M(\tb\,) - m_t \right| + \left|
  M(W) - m_W \right|\,,
\label{eq:q}
\end{equation}
where $M(t)$, $M(\bar{t}\,)$ and $M(W)$ are the reconstructed invariant masses of the top, anti-top quark and $W$ gauge boson respectively. We  identify twelve different resonance histories that are listed in Table \ref{tab:decay_histories_LO}. These  twelve  categories are not sufficient if NLO QCD calculations are considered. If there is an additional resolved light jet, it should be included in the list as well. In practice, to closely mimic what is done on the experimental side the additional light jet is added to the resonance history only if it passes all the cuts. At NLO  a total of 36 different possibilities have to be considered and for each history the ${\cal Q}$ quantity is computed. We note here that the light jet (jets) is present in the resonance histories but its momentum is not taken into account to reconstruct the top-quark momentum even if the decay history indicates that this light jet comes from the top-quark decay.
\begingroup
\begin{table}[h]
    \centering
    \begin{tabular}{ c c c c cc }
    \hline\hline
    &&&&&\\[-0.2cm]
        \textsc{Decay History} & $t$& & $\bar{t}$ & & $W^+$\\
        &&&&&\\[-0.2cm]
         \hline\hline 
%         \toprule
&&&&&\\[-0.2cm]
        1 & $e^+\nu_e \, b_1$ && $\mu^- \bar{\nu}_{\mu}\,b_2$ & &$\tau^+\nu_{\tau}$\\[0.2cm]
       2 &  $e^+\nu_e\,b_2$ && $\mu^- \bar{\nu}_{\mu}\,b_1$ & &$\tau^+\nu_{\tau}$\\[0.2cm]
       3 &  $\tau^+\nu_{\tau}\,b_1$ && $\mu^- \bar{\nu}_{\mu}\,b_2$ & &$e^+\nu_e$\\[0.2cm]
       4 &  $\tau^+\nu_{\tau}\,b_2$ && $\mu^- \bar{\nu}_{\mu}\,b_1$ & &$e^+\nu_e$\\[0.2cm]
      5  &  $e^+\nu_e\,b_1\,j_1$ && $\mu^- \bar{\nu}_{\mu}\,b_2$ & &$\tau^+\nu_{\tau}$\\[0.2cm]
      6  &  $e^+\nu_e\,b_2\,j_1$ && $\mu^- \bar{\nu}_{\mu}\,b_1$ & &$\tau^+\nu_{\tau}$\\[0.2cm]
      7  &  $\tau^+\nu_{\tau}\,b_1\,j_1$ && $\mu^- \bar{\nu}_{\mu}\,b_2$ & &$e^+\nu_e$\\[0.2cm]
     8   &  $\tau^+\nu_{\tau}\,b_2\,j_1$ && $\mu^- \bar{\nu}_{\mu}\,b_1$ & &$e^+\nu_e$\\[0.2cm]
      9  &  $e^+\nu_e\,b_1$ & &$\mu^- \bar{\nu}_{\mu}\,b_2\,j_1$ & &$\tau^+\nu_{\tau}$\\[0.2cm]
      10  &  $e^+\nu_e\,b_2$ && $\mu^- \bar{\nu}_{\mu}\,b_1\,j_1$ & &$\tau^+\nu_{\tau}$\\[0.2cm]
      11  &  $\tau^+\nu_{\tau}\,b_1$ && $\mu^- \bar{\nu}_{\mu}\,b_2\,j_1$ & &$e^+\nu_e$\\[0.2cm]
      12  &  $\tau^+\nu_{\tau}\,b_2$ && $\mu^- \bar{\nu}_{\mu}\,b_1\,j_1$ & &$e^+\nu_e$\\
      &&&&&\\[-0.2cm]
          \hline\hline
    \end{tabular}
    \caption{\it Considered decay histories for $t$, $\bar{t}$ and $W^+$ for events with one resolved light jet.}
    \label{tab:decay_histories_LO}
\end{table}
\endgroup
Jets are constructed out of all final-state partons with pseudo-rapidity $|\eta|<5$ via the  infrared-safe anti-$k_T$ jet algorithm \cite{Cacciari:2008gp} with a clustering parameter $R=0.4$. We require exactly two $b$-jets, at least one light jet and three charged leptons. The two $b$-jets and the light jet of each event need to be located in the central rapidity regions of the detector 
\begin{equation}
|y(b)| < 2.5\,, \quad \quad 
\quad \quad \quad \quad \quad \quad \quad \quad |y(j)| < 2.5 \,,
\end{equation}
and  have a minimum transverse momentum of 
\begin{equation}
p_T(b) > 25~\textrm{GeV}\,, \quad \quad 
\quad \quad \quad \quad \quad \quad \quad \quad
p_T(j) > 25~\textrm{GeV}\,.
\end{equation}
The light jet has to be isolated from  the two $b$-jets according to 
$\Delta R (bj) > 0.4$.   Charged leptons need to fulfill cuts on transverse momenta, rapidities, and be well separated from any jet in the rapidity-azimuthal angle plane as well as from each other 
\begin{equation}
p_T(\ell) > 25~\textrm{GeV}\,, \quad \quad|y(\ell)| < 2.5\,,
\quad \quad \Delta R (b\ell) > 0.4\,, \quad \quad
\Delta R (j\ell) > 0.4\,, \quad \quad  
\Delta R (\ell\ell) > 0.4\,,
\end{equation}
where $\ell = e^+, \tau^+, \mu^-$. There are no restrictions on the kinematics of the second light jet (if resolved) and the missing transverse momentum.

% =============================================
%
\section{Theoretical results for the integrated fiducial cross section}
\label{sec:fiducialxs}
%
% =============================================

%============================================
\begin{figure}[t!]
    \centering
    \includegraphics[width=0.8\textwidth]{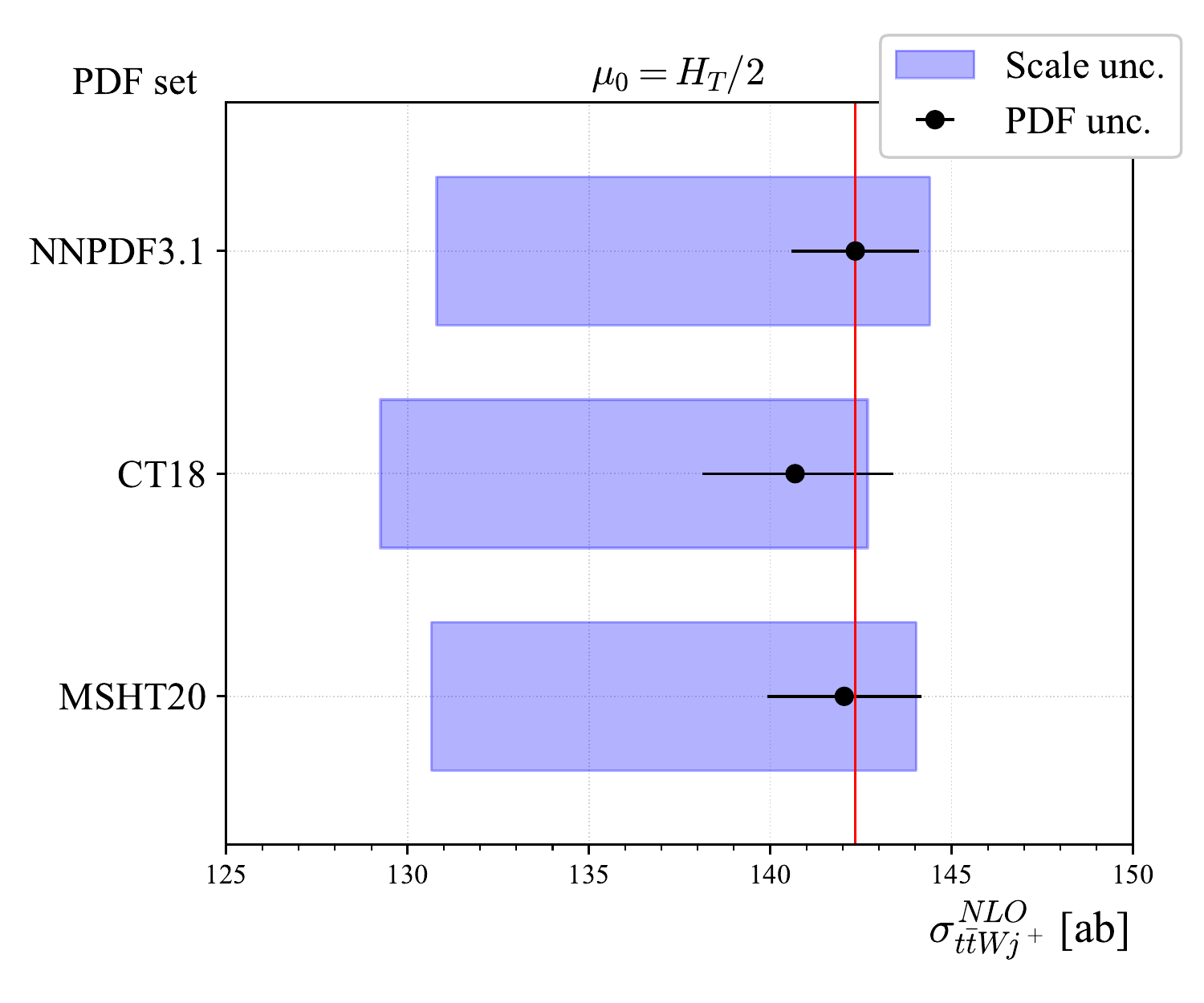}
    \caption{\it 
    Integrated fiducial cross sections at NLO in QCD for the $pp \to \WWW \, j +X$ process at the LHC with $\sqrt{s}=13$ TeV. Theoretical results are provided for $\mu_R = \mu_F = \mu_0 = H_T/2$ as well as for the following NLO PDF sets: NNPDF3.1, CT18 and MSHT20. Scale uncertainties are reported for each case  together with  internal PDF uncertainties. The red line represents the central prediction of our default setup.}
    \label{fig:fiducialxs_pdf_uncertainties}
\end{figure}
%============================================
\begin{figure}[t]
  \begin{center}
    \includegraphics[width=0.8\textwidth]{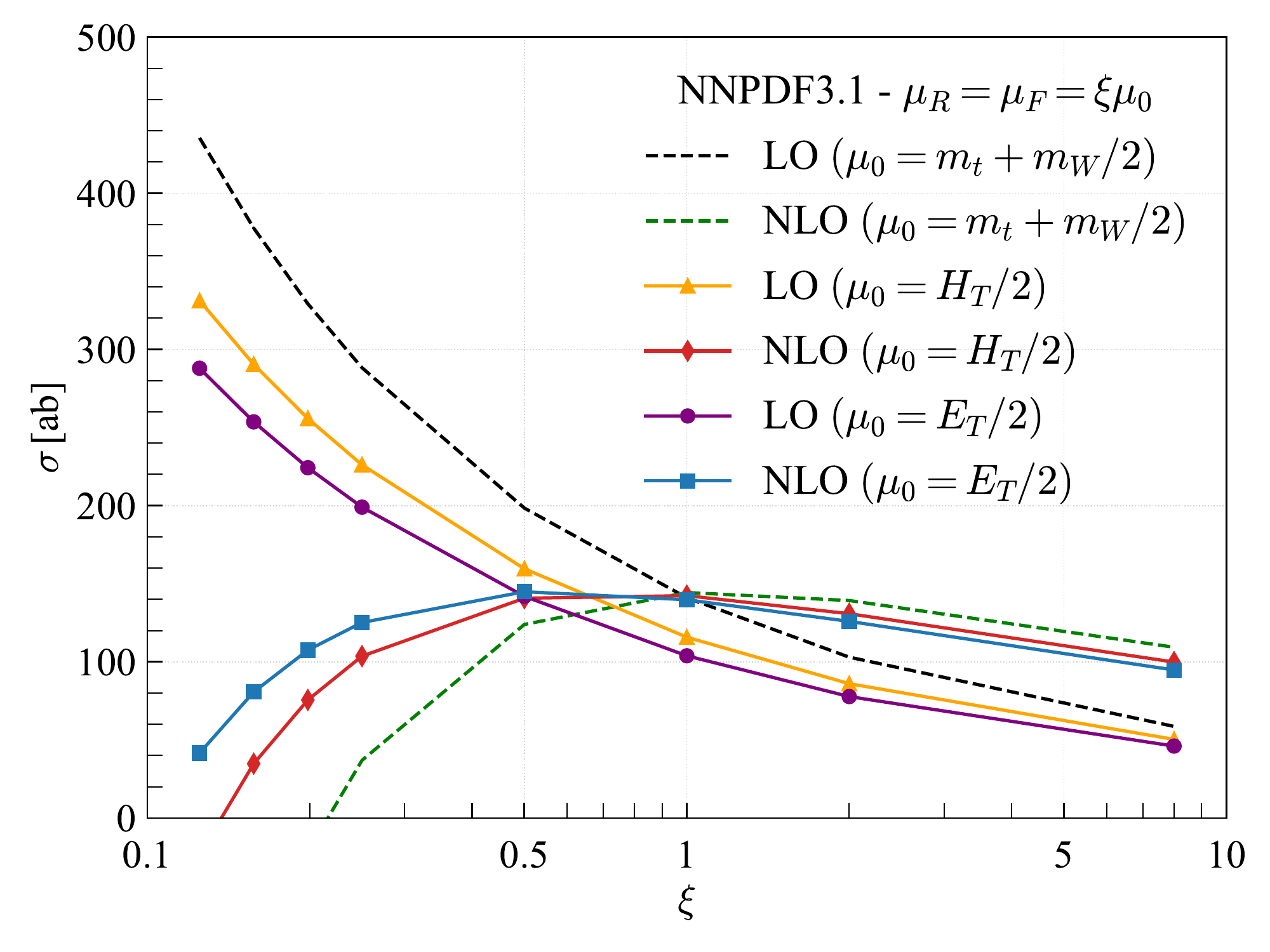}
\end{center}
\caption{\label{fig:scaledep_muF+muR} \it 
Scale dependence of the integrated fiducial cross section at LO and NLO in QCD for the 
$pp \to \WWW \, j+X$ process at the LHC with $\sqrt{s}= 13$ TeV.
Renormalisation and factorisation scales are set to the common value
$\mu_R=\mu_F=\xi\mu_0$ where $\mu_0=H_T/2$, $\mu_0=m_t+m_W/2$ and $\mu_0=E_T/2$.
 (N)LO  NNPDF3.1 PDFs are employed.}
 \end{figure}
%=============================================
\begin{figure}[t]
  \begin{center}
    \includegraphics[width=.49\textwidth]{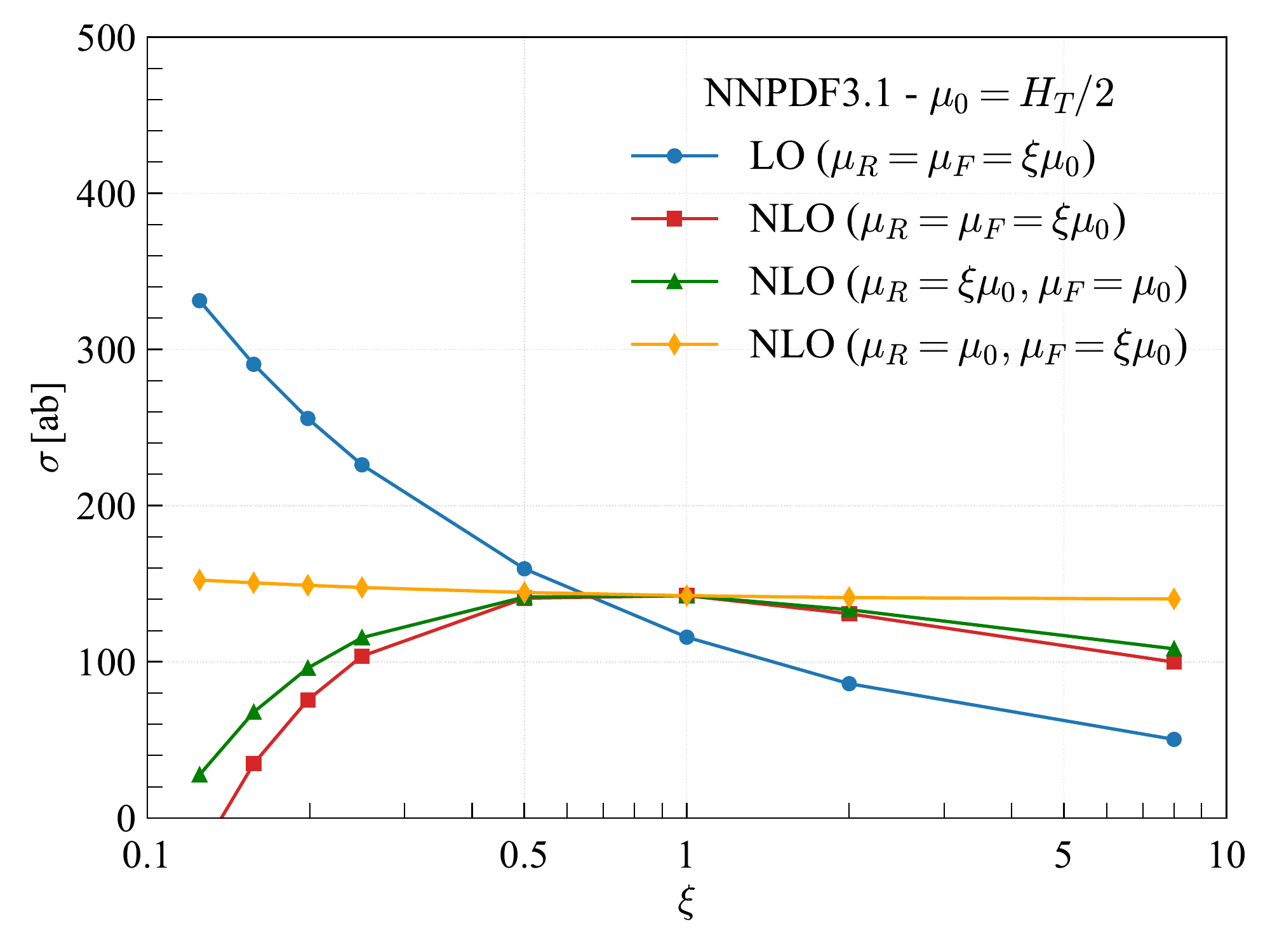} 
    \includegraphics[width=.49\textwidth]{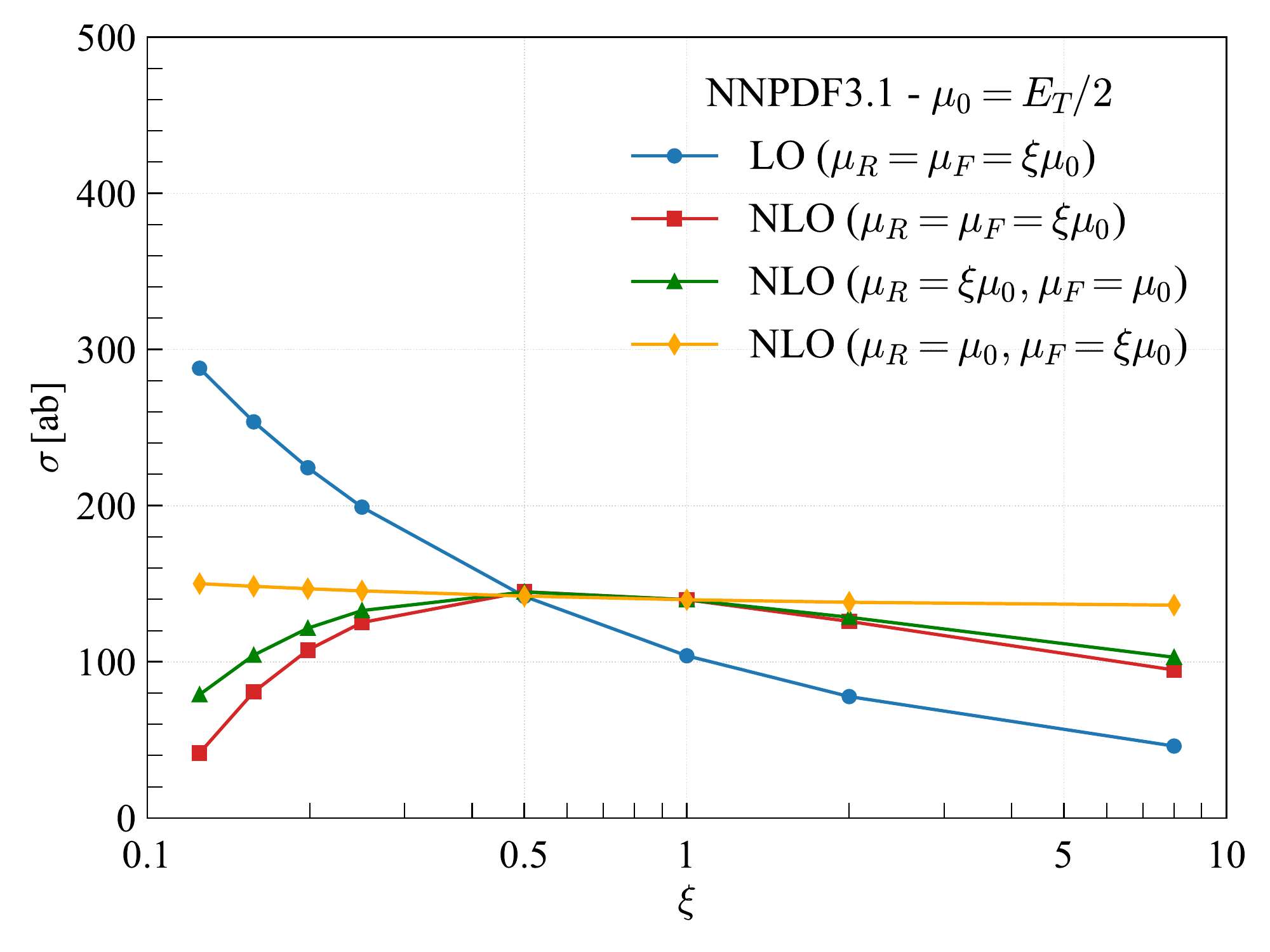}
    \includegraphics[width=.49\textwidth]{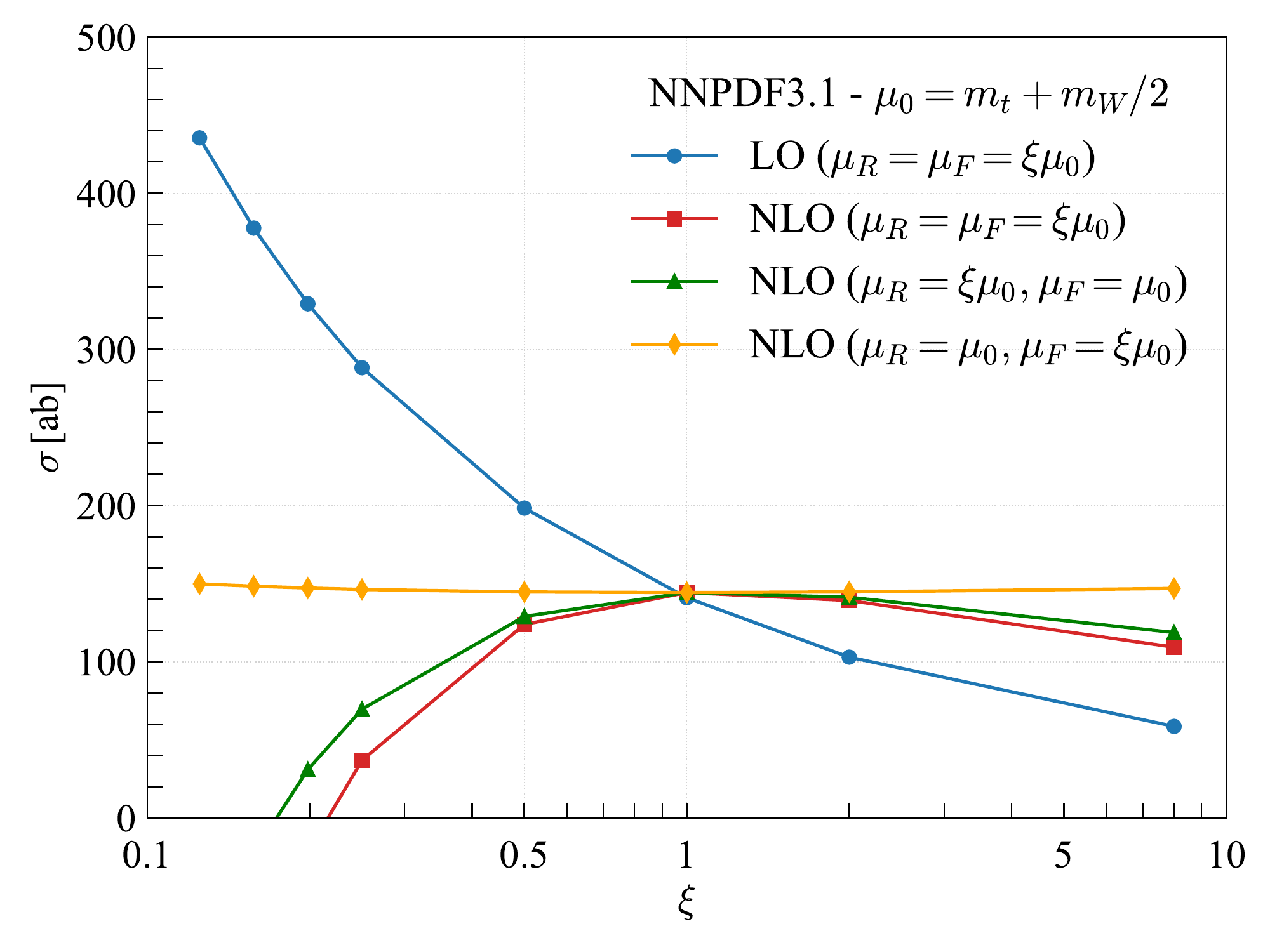}
\end{center}
\caption{\label{fig:scaledep_muFvsmuR_HTET} \it 
Scale dependence of the integrated fiducial cross section at LO and NLO in QCD for the 
$pp \to \WWW \, j+X$ process at the LHC with $\sqrt{s}= 13$ TeV. Renormalisation and factorisation scales are set to the common value $\mu_R=\mu_F=\xi\mu_0$ where $\mu_0=H_T/2$, $\mu_0=m_t+m_W/2$ and $\mu_0=E_T/2$.  (N)LO NNPDF3.1 PDFs are employed. For each case of $\mu_0$ also shown is the variation of $\mu_R$ with fixed $\mu_F$ and the variation of $\mu_F$ with fixed $\mu_R$.}
 \end{figure}
%=============================================

With the Standard Model parameters and fiducial phase-space cuts specified in Section \ref{sec:inputparameters}, we arrive at the following predictions for the $pp \to \WWW j + X$ production process at the LHC with
 $\sqrt{s}=13$ TeV 
\begin{equation}
\begin{split}
  \sigma^{\rm LO}_{pp \to t\tb W^+j}(\textrm{NNPDF3.1}, \mu_0=H_T/2)  
  & = 115.8^{+38\%}_{-26\%} \, [\textrm{scale}]\,\textrm{ab} \,,\\[0.2cm]
  \sigma^{\rm NLO}_{pp \to t\tb W^+j}(\textrm{NNPDF3.1}, \mu_0=H_T/2)  
  & = 142.3^{+1.4\%}_{-8.1\%}\, [\textrm{scale}] \, {}^{+1.2\%}_{-1.2\%} \,[\textrm{PDF}]\,\textrm{ab} \,,
  \end{split}
\end{equation}
where LO and NLO NNPDF3.1 PDF sets and our default  $\mu_0=\mu_R=\mu_F=H_T/2$ scale setting have been employed. At the central value of the scale   $\mu_0 = H_T/2$, the $q\bar{q}^{\, \prime}$ channel dominates the LO $pp$ cross section by $62\%$, followed by the $gq$ channel with $38\%$. We can compare this to the on-shell $pp\to t\bar{t}W^+ j$ production process with stable top quarks and $W$ gauge boson for $p_T(j)> 25$ GeV,  where both $q\bar{q}^{\, \prime}$ and $gq$ are equally important.  The change in the size of individual contributions depends, of course, on the fiducial regions required in the particular analysis. Indeed, only when the transverse momentum cut on the light jet is increased to $50$ GeV in our off-shell calculations we obtain the same result as for the stable $pp \to t\bar{t}W^+j$ process. The latter process, however, is already dominated by the $gq$ channel $(60\%)$ for $p_T(j)> 50$ GeV cut.  This, of course, tells us that we should not expect similar results in the fiducial phase-space regions and for the full phase space. If the importance of the sub-processes already depends so strongly on the used cuts, the same may be true for the size of the NLO corrections or the theoretical uncertainties associated with the NLO cross section. In the case at hand, the full $pp$ cross section receives moderate and positive NLO corrections of $23\%$. The theoretical uncertainties resulting from scale variations taken as a maximum of the lower and upper bounds are $38\%$ at LO and $8\%$ at NLO.  The internal NNPDF3.1 PDF uncertainties, which are another source of theoretical uncertainties, are of the order of $1\%$. They are, therefore, well below the theoretical uncertainties due to the scale dependence. The latter remains the dominant source of the theoretical systematics.  However, for similar processes differences coming from NLO results for various PDF sets can be  higher than the individual estimates of PDF systematics, see e.g. Refs. \cite{Bevilacqua:2016jfk,Bevilacqua:2021cit,Stremmer:2021bnk,Bevilacqua:2022nrm}. We checked whether this is the case with the current process.  Our findings for CT18/CT14 and MSHT20 PDF sets can be summarised as follows:
\begin{equation}
\begin{split}
\sigma^{\rm LO}_{pp \to t\tb W^+j}(\textrm{CT14}, \mu_0=H_T/2)  &=
  133.1^{+41\%}_{-27\%}\, [\textrm{scale}]\,\textrm{ab}\,,\\[0.2cm]
  \sigma^{\rm NLO}_{pp \to t\tb W^+j}(\textrm{CT18}, \mu_0=H_T/2)  &=
 140.7^{+1.4\%}_{-8.1\%}\, [\textrm{scale}]\, {}^{+1.9\%}_{-1.8\%} \, 
 [\textrm{PDF}]\,\textrm{ab}\,,\\[0.2cm]
  \sigma^{\rm LO}_{pp \to t\tb W^+j}(\textrm{MSHT20}, \mu_0=H_T/2)  &=
   124.5^{+42\%}_{-27\%}\, [\textrm{scale}]\,\textrm{ab}\,,\\[0.2cm]
  \sigma^{\rm NLO}_{pp \to t\tb W^+j}(\textrm{MSHT20}, \mu_0=H_T/2) & =
   142.0^{+1.4\%}_{-8.0\%} \,[\textrm{scale}]\, {}^{+1.5\%}_{-1.5\%} \, [\textrm{PDF}]\,\textrm{ab} \,.
\end{split}
\end{equation}
The three PDF choices yield very consistent NLO results with differences of up to $1\%$ at most. The latter differences are consistent with the obtained internal PDF  uncertainties, which are maximally at the level of $2\%$.  In addition, the size of the theoretical uncertainties arising from scale variations is the same as for NNPDF3.1, CT18 and MSHT20. The only difference that we can observe is in the magnitude of higher-order effects. The latter, however, are driven by the LO cross section and  the associated value of $\alpha_s(m_Z)$ used in the PDFs.  Consequently, for the CT family we obtain the following ${\cal K}$-factor: ${\cal K}= \sigma^{\rm NLO}_{pp \to t\tb W^+j}/\sigma^{\rm LO}_{pp \to t\tb W^+j}=1.06$, while for  MSHT20  we have instead ${\cal K}=1.14$.  A graphical representation of these findings is given in Figure \ref{fig:fiducialxs_pdf_uncertainties} where we show the NLO integrated fiducial cross sections for the three main families of PDFs separately, together with the corresponding scale and PDF uncertainties.
%
%=============================================
\begin{table}[t]
\begin{center}
\begin{tabular}{cccccccc}
  \hline \hline
  &&&&&&&\\[-0.2cm]
  PDF &$p_{T}(b)$ & $\sigma^{\textrm{LO}}$  [ab]
      & $\delta_{scale}$ & $\sigma^{\textrm{NLO}}$ [ab]
      & $\delta_{scale}$ & $\delta_{\textrm{PDF}}$ & $\mathcal{K}$ \\
  &&&&&&&\\[-0.2cm]
  \hline
  \hline
  &&&&&&&\\[-0.2cm]
NNPDF & 25 & 115.8 & $^{ +43.8 ~(38 \%) }_{ -29.8 ~(26 \%) }$ 
& 142.3 & $^{ +~2.1 ~(1.4 \%) }_{ -11.5 ~(8.1 \%) }$ 
& $^{ +1.8 ~(1.2 \%) }_{ -1.8 ~(1.2 \%) }$ & 1.23\\[0.2cm]
      & 30 & 106.1 & $^{ +40.2 ~(38 \%) }_{ -27.3 ~(26 \%) }$ 
      & 129.8 & $^{ +~1.9 ~(1.4 \%) }_{ -10.4 ~(8.0 \%) }$ 
      & $^{ +1.6 ~(1.2 \%) }_{ -1.6 ~(1.2 \%) }$ & 1.22\\[0.2cm]
      & 35 &  96.1 & $^{ +36.4 ~(38 \%) }_{ -24.7 ~(26 \%) }$ 
      & 117.3 & $^{ +~1.7 ~(1.4 \%) }_{  -~9.4 ~(8.0 \%) }$ 
      & $^{ +1.4 ~(1.2 \%) }_{ -1.4 ~(1.2 \%) }$ & 1.22\\[0.2cm]
      & 40 &  86.1 & $^{ +32.7 ~(38 \%) }_{ -22.2 ~(26 \%) }$ 
      & 105.0 & $^{ +~1.5 ~(1.5 \%) }_{  -~8.4 ~(8.0 \%) }$ 
      & $^{ +1.3 ~(1.2 \%) }_{ -1.3 ~(1.2 \%) }$ & 1.22\\
  &&&&&&&\\[-0.2cm]
  \hline 
  \hline
  &&&&&&&\\[-0.2cm]
CT   & 25 & 133.1 & $^{ +54.8 ~(41 \%) }_{ -36.2 ~(27 \%) }$ 
& 140.7 & $^{ +~2.0 ~(1.4 \%) }_{ -11.4 ~(8.1 \%) }$ 
& $^{ +2.7 ~(1.9 \%) }_{ -2.6 ~(1.8 \%) }$ & 1.06\\[0.2cm]
     & 30 & 122.0 & $^{ +50.2 ~(41 \%) }_{ -33.2 ~(27 \%) }$ 
     & 128.3 & $^{ +~1.8 ~(1.4 \%) }_{ -10.3 ~(8.0 \%) }$ 
     & $^{ +2.5 ~(1.9 \%) }_{ -2.3 ~(1.8 \%) }$ & 1.05\\[0.2cm]
     & 35 & 110.4 & $^{ +45.5 ~(41 \%) }_{ -30.1 ~(27 \%) }$ 
     & 115.9 & $^{ +~1.6 ~(1.4 \%) }_{  -~9.3 ~(8.0 \%) }$ 
     & $^{ +2.2 ~(1.9 \%) }_{ -2.1 ~(1.8 \%) }$ & 1.05\\[0.2cm]
     & 40 &  99.0 & $^{ +40.8 ~(41 \%) }_{ -27.0 ~(27 \%) }$ 
     & 103.8 & $^{ +~1.5 ~(1.4 \%) }_{  -~8.3 ~(8.0 \%) }$ 
     & $^{ +2.0 ~(1.9 \%) }_{ -1.9 ~(1.8 \%) }$ & 1.05\\
  &&&&&&&\\[-0.2cm]
  \hline
  \hline
  &&&&&&&\\[-0.2cm]
MSHT   & 25 & 124.5 & $^{ +51.7 ~(42 \%) }_{ -34.1 ~(27 \%) }$ 
& 142.0 & $^{ +~2.0 ~(1.4 \%) }_{ -11.4 ~(8.0 \%) }$ 
& $^{ +2.1 ~(1.5 \%) }_{ -2.1 ~(1.5 \%) }$ & 1.14\\[0.2cm]
       & 30 & 114.0 & $^{ +47.4 ~(42 \%) }_{ -31.2 ~(27 \%) }$ 
       & 129.6 & $^{ +~1.8 ~(1.4 \%) }_{ -10.3 ~(7.9 \%) }$ 
       & $^{ +1.9 ~(1.5 \%) }_{ -1.9 ~(1.5 \%) }$ & 1.14\\[0.2cm]
       & 35 & 103.2 & $^{ +42.9 ~(42 \%) }_{ -28.3 ~(27 \%) }$ 
       & 117.0 & $^{ +~1.6 ~(1.4 \%) }_{  -~9.3 ~(7.9 \%) }$ 
       & $^{ +1.8 ~(1.5 \%) }_{ -1.7 ~(1.5 \%) }$ & 1.13\\[0.2cm]
       & 40 &  92.5 & $^{ +38.5 ~(42 \%) }_{ -25.4 ~(27 \%) }$ 
       & 104.9 & $^{ +~1.5 ~(1.4 \%) }_{  -~8.3 ~(7.9 \%) }$ 
       & $^{ +1.6 ~(1.5 \%) }_{ -1.6 ~(1.5 \%) }$ & 1.13\\
  &&&&&&&\\[-0.2cm]
  \hline
  \hline
\end{tabular}
\caption{\it Integrated fiducial cross sections at LO and NLO in QCD for the 
$pp \to \WWW \, j+X$ process at the LHC with $\sqrt{s}= 13$ TeV.  Results 
for a different value of the cut on the transverse momentum of the $b$-jet, $p_T(b)$, are presented. The theoretical uncertainties coming from the 7-point scale variation are denoted as $\delta_{scale}$, whereas the internal PDF uncertainties are labeled as $\delta_{\rm PDF}$. Results are shown for $\mu_0 = H_T/2$ and for the following (N)LO PDF sets: NNPDF3.1, CT14/CT18 as well as for MSHT20. In the last column the ${\cal K}$-factor is also given.}
\label{table:stability_ptb}
\end{center}
\end{table}
%=============================================
\begin{table}[t]
\begin{center}
\begin{tabular}{cccccccc}
  \hline \hline
  &&&&&&&\\[-0.2cm]
  PDF &$p_{T}(j)$ & $\sigma^{\textrm{LO}}$  [ab]
      & $\delta_{scale}$ & $\sigma^{\textrm{NLO}}$ [ab]
      & $\delta_{scale}$ & $\delta_{\textrm{PDF}}$ & $\mathcal{K}$ \\
  &&&&&&&\\[-0.2cm]
  \hline
  \hline
  &&&&&&&\\[-0.2cm]
NNPDF   & 25 & 115.8 & $^{ +43.8 ~(38 \%) }_{ -29.8 ~(26 \%) }$ 
& 142.3 & $^{ +~2.1 ~(1.4 \%) }_{ -11.5 ~(8.1 \%) }$ 
& $^{ +1.8 ~(1.2 \%) }_{ -1.8 ~(1.2 \%) }$ & 1.23\\[0.2cm]
        & 30 & 103.4 & $^{ +39.4 ~(38 \%) }_{ -26.7 ~(26 \%) }$ 
        & 130.8 & $^{ +~2.1 ~(1.6 \%) }_{ -11.5 ~(8.8 \%) }$ 
        & $^{ +1.6 ~(1.2 \%) }_{ -1.6 ~(1.2 \%) }$ & 1.27\\[0.2cm]
        & 35 &  93.5 & $^{ +35.8 ~(38 \%) }_{ -24.3 ~(26 \%) }$ 
        & 121.2 & $^{ +~2.1 ~(1.7 \%) }_{ -11.4 ~(9.4 \%) }$ 
        & $^{ +1.4 ~(1.1 \%) }_{ -1.4 ~(1.1 \%) }$ & 1.30\\[0.2cm]
        & 40 &  85.4 & $^{ +32.9 ~(38 \%) }_{ -22.2 ~(26 \%) }$ 
        & 112.9 & $^{ +~2.7 ~(2.4 \%) }_{ -11.1 ~(9.8 \%) }$ 
        & $^{ +1.3 ~(1.1 \%) }_{ -1.3 ~(1.1 \%) }$ & 1.32\\
  &&&&&&&\\[-0.2cm]
  \hline 
  \hline
  &&&&&&&\\[-0.2cm]
CT   & 25 & 133.1 & $^{ +54.8 ~(41 \%) }_{ -36.2 ~(27 \%) }$ 
& 140.7 & $^{ +~2.0 ~(1.4 \%) }_{ -11.4 ~(8.1 \%) }$ 
& $^{ +2.7 ~(1.9 \%) }_{ -2.6 ~(1.8 \%) }$ & 1.06\\[0.2cm]
     & 30 & 119.2 & $^{ +49.3 ~(41 \%) }_{ -32.6 ~(27 \%) }$ 
     & 129.4 & $^{ +~2.0 ~(1.6 \%) }_{ -11.4 ~(8.8 \%) }$ 
     & $^{ +2.5 ~(1.9 \%) }_{ -2.3 ~(1.8 \%) }$ & 1.08\\[0.2cm]
     & 35 & 108.2 & $^{ +45.0 ~(42 \%) }_{ -29.7 ~(27 \%) }$ 
     & 119.9 & $^{ +~2.1 ~(1.7 \%) }_{ -11.2 ~(9.4 \%) }$ 
     & $^{ +2.3 ~(1.9 \%) }_{ -2.1 ~(1.7 \%) }$ & 1.11\\[0.2cm]
     & 40 &  99.1 & $^{ +41.4 ~(42 \%) }_{ -27.3 ~(28 \%) }$ 
     & 111.7 & $^{ +~2.8 ~(2.5 \%) }_{ -11.0 ~(9.8 \%) }$ 
     & $^{ +2.1 ~(1.9 \%) }_{ -1.9 ~(1.7 \%) }$ & 1.13\\
  &&&&&&&\\ [-0.2cm] \hline
  \hline
  &&&&&&&\\[-0.2cm]
MSHT   & 25 & 124.5 & $^{ +51.7 ~(42 \%) }_{ -34.1 ~(27 \%) }$ 
& 142.0 & $^{ +~2.0 ~(1.4 \%) }_{ -11.4 ~(8.0 \%) }$ 
& $^{ +2.1 ~(1.5 \%) }_{ -2.1 ~(1.5 \%) }$ & 1.14\\[0.2cm]
       & 30 & 111.4 & $^{ +46.6 ~(42 \%) }_{ -30.7 ~(28 \%) }$ 
       & 130.6 & $^{ +~2.0 ~(1.6 \%) }_{ -11.4 ~(8.7 \%) }$ 
       & $^{ +1.9 ~(1.5 \%) }_{ -1.9 ~(1.4 \%) }$ & 1.17\\[0.2cm]
       & 35 & 101.0 & $^{ +42.5 ~(42 \%) }_{ -27.9 ~(28 \%) }$ 
       & 121.0 & $^{ +~2.1 ~(1.7 \%) }_{ -11.2 ~(9.3 \%) }$ 
       & $^{ +1.7 ~(1.4 \%) }_{ -1.7 ~(1.4 \%) }$ & 1.20\\[0.2cm]
       & 40 &  92.5 & $^{ +39.1 ~(42 \%) }_{ -25.7 ~(28 \%) }$ 
       & 112.8 & $^{ +~2.6 ~(2.3 \%) }_{ -11.0 ~(9.7 \%) }$ 
       & $^{ +1.6 ~(1.4 \%) }_{ -1.5 ~(1.3 \%) }$ & 1.22\\
  &&&&&&&\\[-0.2cm]
  \hline
  \hline
\end{tabular}
\caption{\it 
  Integrated fiducial cross sections at LO and NLO in QCD for the 
$pp \to \WWW \, j+X$ process at the LHC with $\sqrt{s}= 13$ TeV.  Results for a different value of the cut on the transverse momentum of the light jet, $p_T(j)$, are presented. The theoretical uncertainties coming from 
the 7-point scale variation are denoted as $\delta_{scale}$, whereas the internal PDF uncertainties are labeled as $\delta_{\rm PDF}$. Results are shown for $\mu_0 = H_T/2$ and for the following (N)LO PDF sets: NNPDF3.1, CT14/CT18 as well as for MSHT20. In the last column the ${\cal K}$-factor is also given. }
\label{table:stability_ptj}
\end{center}
\end{table}
%=============================================
\begin{table}[t]
    \begin{center}
    \begin{tabular}{cccccccc}
        \hline
        \hline
        &&&&&\\[-0.2cm]
      PDF & $p_T^{miss}\,$[GeV] & $\sigma^{\rm LO}\,$[ab]
            & $\delta_{scale}$ &$ \sigma^{\rm NLO}\,$[ab] 
            & $\delta_{scale}$
            & $\delta_{\textrm{PDF}}$ & $\mathcal{K}$ \\
        &&&&&\\[-0.2cm]
        \hline
        \hline
        &&&&&\\[-0.2cm]
NNPDF &  0 & 115.8 & $^{ +43.8 ~(38 \%) }_{ -29.8 ~(26 \%) }$ 
& 142.3 & $^{ +~2.1 ~(1.4 \%) }_{ -11.5 ~(8.1 \%) }$ 
& $^{ +1.8 ~(1.2 \%) }_{ -1.8 ~(1.2 \%) }$ & 1.23\\[0.2cm]
      & 10 & 114.6 & $^{ +43.4 ~(38 \%) }_{ -29.5 ~(26 \%) }$ 
      & 140.9 & $^{ +~2.0 ~(1.4 \%) }_{ -11.4 ~(8.1 \%) }$ 
      & $^{ +1.7 ~(1.2 \%) }_{ -1.7 ~(1.2 \%) }$ & 1.23\\[0.2cm]
      & 20 & 111.2 & $^{ +42.1 ~(38 \%) }_{ -28.6 ~(26 \%) }$ 
      & 137.1 & $^{ +~2.0 ~(1.5 \%) }_{ -11.2 ~(8.2 \%) }$ 
      & $^{ +1.7 ~(1.2 \%) }_{ -1.7 ~(1.2 \%) }$ & 1.23\\[0.2cm]
      & 30 & 105.9 & $^{ +40.1 ~(38 \%) }_{ -27.2 ~(26 \%) }$ 
      & 131.0 & $^{ +~1.9 ~(1.5 \%) }_{ -10.8 ~(8.3 \%) }$ 
      & $^{ +1.6 ~(1.2 \%) }_{ -1.6 ~(1.2 \%) }$ & 1.24\\[0.2cm]
      & 40 &  99.0 & $^{ +37.5 ~(38 \%) }_{ -25.5 ~(26 \%) }$ 
      & 122.8 & $^{ +~1.9 ~(1.5 \%) }_{ -10.3 ~(8.4 \%) }$ 
      & $^{ +1.5 ~(1.2 \%) }_{ -1.5 ~(1.2 \%) }$ & 1.24\\
          &&&&&\\[-0.2cm]
         \hline
         \hline
         &&&&&\\[-0.2cm]
CT & 0  & 133.1 & $^{ +54.8 ~(41 \%) }_{ -36.2 ~(27 \%) }$ 
& 140.7 & $^{ +~2.0 ~(1.4 \%) }_{ -11.4 ~(8.1 \%) }$ 
& $^{ +2.7 ~(1.9 \%) }_{ -2.6 ~(1.8 \%) }$ & 1.06\\[0.2cm]
   & 10 & 131.7 & $^{ +54.2 ~(41 \%) }_{ -35.8 ~(27 \%) }$ 
   & 139.3 & $^{ +~2.0 ~(1.4 \%) }_{ -11.3 ~(8.1 \%) }$ 
   & $^{ +2.7 ~(1.9 \%) }_{ -2.5 ~(1.8 \%) }$ & 1.06\\[0.2cm]
   & 20 & 127.8 & $^{ +52.6 ~(41 \%) }_{ -34.8 ~(27 \%) }$ 
   & 135.5 & $^{ +~1.9 ~(1.4 \%) }_{ -11.1 ~(8.2 \%) }$ 
   & $^{ +2.6 ~(1.9 \%) }_{ -2.5 ~(1.8 \%) }$ & 1.06\\[0.2cm]
   & 30 & 121.7 & $^{ +50.1 ~(41 \%) }_{ -33.1 ~(27 \%) }$ 
   & 129.4 & $^{ +~1.9 ~(1.5 \%) }_{ -10.7 ~(8.3 \%) }$ 
   & $^{ +2.5 ~(1.9 \%) }_{ -2.4 ~(1.8 \%) }$ & 1.06\\[0.2cm]
   & 40 & 113.8 & $^{ +46.9 ~(41 \%) }_{ -31.0 ~(27 \%) }$ 
   & 121.4 & $^{ +~1.8 ~(1.5 \%) }_{ -10.2 ~(8.4 \%) }$ 
   & $^{ +2.4 ~(1.9 \%) }_{ -2.2 ~(1.8 \%) }$ & 1.07\\
         &&&&&\\[-0.2cm] 
         \hline
         \hline
         &&&&&\\[-0.2cm]
MSHT  & 0  & 124.5 & $^{ +51.7 ~(42 \%) }_{ -34.1 ~(27 \%) }$ 
& 142.0 & $^{ +~2.0 ~(1.4 \%) }_{ -11.4 ~(8.0 \%) }$ 
& $^{ +2.1 ~(1.5 \%) }_{ -2.1 ~(1.5 \%) }$ & 1.14\\[0.2cm]
      & 10 & 123.2 & $^{ +51.2 ~(42 \%) }_{ -33.7 ~(27 \%) }$ 
      & 140.6 & $^{ +~2.0 ~(1.4 \%) }_{ -11.3 ~(8.0 \%) }$ 
      & $^{ +2.1 ~(1.5 \%) }_{ -2.1 ~(1.5 \%) }$ & 1.14\\[0.2cm]
      & 20 & 119.5 & $^{ +49.7 ~(42 \%) }_{ -32.7 ~(27 \%) }$ 
      & 136.8 & $^{ +~1.9 ~(1.4 \%) }_{ -11.0 ~(8.1 \%) }$ 
      & $^{ +2.1 ~(1.5 \%) }_{ -2.0 ~(1.5 \%) }$ & 1.14\\[0.2cm]
      & 30 & 113.8 & $^{ +47.3 ~(42 \%) }_{ -31.2 ~(27 \%) }$ 
      & 130.7 & $^{ +~1.9 ~(1.4 \%) }_{ -10.7 ~(8.2 \%) }$ 
      & $^{ +2.0 ~(1.5 \%) }_{ -1.9 ~(1.5 \%) }$ & 1.15\\[0.2cm]
      & 40 & 106.3 & $^{ +44.2 ~(42 \%) }_{ -29.1 ~(27 \%) }$ 
      & 122.6 & $^{ +~1.8 ~(1.5 \%) }_{ -10.2 ~(8.3 \%) }$ 
      & $^{ +1.8 ~(1.5 \%) }_{ -1.8 ~(1.5 \%) }$ & 1.15\\
        &&&&&\\[-0.2cm]    
        \hline
        \hline                                      
    \end{tabular}
    \caption{\it
      Integrated fiducial cross sections at LO and NLO in QCD for the 
$pp \to \WWW \, j+X$ process at the LHC with $\sqrt{s}= 13$ TeV.  Results for a different value of the cut on the missing transverse momentum, $p_T^{miss}$, are presented. The theoretical uncertainties coming from 
the 7-point scale variation are denoted as $\delta_{scale}$, whereas the internal PDF uncertainties are labeled as $\delta_{\rm PDF}$. Results are shown for $\mu_0 = H_T/2$ and for the following (N)LO PDF sets: NNPDF3.1, CT14/CT18 as well as for MSHT20. In the last column the ${\cal K}$-factor is also given.}
    \label{table:stability_ptmiss}
    \end{center}
\end{table}
%=============================================

In addition to our default scale choice, $\mu_0=H_T/2$, we also show
results for a fixed scale choice, $\mu_0=m_t+m_W/2$, and a second
dynamical scale setting, $\mu_0=E_T/2$.
\begin{equation}
\begin{split}
  \sigma^{\rm NLO}_{pp \to t\tb W^+j}(\textrm{NNPDF3.1}, \mu_0=m_t+m_W/2) & =
   144.3^{\,+0.3\%}_{-14.1\%}\, [\textrm{scale}]\, {}^{+1.2\%}_{-1.2\%} \,[\textrm{PDF}]\,\textrm{ab}\\[0.2cm]
  \sigma^{\rm NLO}_{pp \to t\tb W^+j}(\textrm{NNPDF3.1}, \mu_0=E_T/2)  
  & = 139.7^{ +3.7 \% }_{ -9.9 \% }\, [\textrm{scale}]\, {}^{+1.2\%}_{-1.2\%} \, [\textrm{PDF}]\,\textrm{ab}
\end{split}
\end{equation}
The NLO QCD results for the three scale settings are very consistent, as the differences that can be observed are at most $3\%$. However, the resulting 
scale uncertainties are larger for the two additional scale settings. Indeed, they increase only slightly for $\mu_0=E_T/2$ and are up to $10\%$, but for $\mu_0=m_t+m_W/2$ they are already at the level of $14\%$.  It is also instructive to present the scale dependence of our results in a more graphical fashion. To this end, we show in Figure \ref{fig:scaledep_muF+muR} the integrated fiducial cross sections at LO and NLO, based on the NNPDF3.1 PDF set. The scales $\mu_R$ and $\mu_F$ are varied simultaneously according to the prescription $\mu_R=\mu_F=\xi \mu_0$, where $\xi \in (0.125, \dots, 8)$, where $\mu_0=H_T/2$, $\mu_0=m_t+m_W/2$ and $\mu_0=E_T/2$.  We can clearly see that in the interval of interest,  i.e. for $\xi \in (0.5, \dots, 2)$,  both dynamic scale settings behave very similarly, while a larger variation can be observed for the fixed scale choice.  For the sake of completeness, in Figure \ref{fig:scaledep_muFvsmuR_HTET} we present afresh the scale dependence of the LO and NLO integrated cross sections for each case of $\mu_0$ separately. Also shown is the variation of $\mu_R$ with the fixed value of $\mu_F$ and the variation of $\mu_F$ with fixed $\mu_R$. In the range $\xi \in (0.5, \dots, 2)$ for the three scale settings the scale variation is driven by the changes in $\mu_R$. In other words, if we changed $\mu_R$ and $\mu_F$ up and down by a factor of $2$ around $\mu_0$ at the same time instead of independently, the scale uncertainties would not change significantly. Not only the choice of $\mu_0$ but also the value of $\xi$ is very important. If $\xi$ is not properly selected it can introduce large higher-order effects and even results in negative NLO cross sections as can be also seen in Figure \ref{fig:scaledep_muFvsmuR_HTET}.

Even though we are reporting results for a rather inclusive fiducial phase-space region, it is very important to investigate the stability of the presented results with respect to small changes introduced into this phase space. To this end, in Table \ref{table:stability_ptb} we provide integrated fiducial cross sections at LO and NLO with a different cut on the transverse momentum of the $b$-jet. Theoretical uncertainties coming from scale variation denoted as $\delta_{scale}$ are also shown together with internal PDF uncertainties, that are labeled as $\delta_{\rm PDF}$. We display results for the three PDF sets. In addition, we provide the ${\cal K}$-factor for each case. The results are evaluated using $\mu_0=H_T/2$. We observe a surprisingly  stable behaviour of the theoretical systematics when varying the $p_T(b)$ cut in the range of $p_T(b) \in (25-40)$ GeV. There is also no change in the relative size of the NLO QCD corrections. When the transverse momentum cut of the light jet is varied in the range of $p_T(j)\in (25-40)$ GeV instead,  the situation is rather similar, as can be seen in Table \ref{table:stability_ptj}. For our default PDF set scale uncertainties increase only slightly from $8\%$ up to $10\%$ and the size of NLO QCD corrections increases from $23\%$ to $32\%$.  The latter effect, however, is due to much larger changes in the LO cross section. Similar findings are obtained for CT18 and MSHT20, showing very good agreement between the results.  Finally, in Table \ref{table:stability_ptmiss} we provide the results for  an analogous stability study, however, this time the missing transverse momentum cut is varied in the range of $p_T^{miss}\in (0-40)$ GeV. In our default setup, we impose no cut on $p_T^{miss}$ so it is interesting to see whether  introducing the $p_T^{miss}$ cut can affect our findings. The $p_T^{miss}$ cut at the level of $30$ GeV or $40$ GeV is very often imposed in ATLAS and CMS  measurements for the $pp \to t\bar{t}W^\pm$ process.   As in the case of the $p_T(b)$ and $p_T(j)$ cut also for the $p_T^{miss}$  one we observe very stable results when increasing its value.

To summarise this part, our higher-order theoretical predictions show a very stable behaviour with respect to theoretical uncertainties. In particular, no large differences can be observed between the results obtained for the highest value of the $p_T(b)$ and $p_T(j)$  cut as well as for the default value of $25$ GeV. This suggests that the perturbative expansion for the $pp \to \WWW \, j+X$  process is not spoiled by the appearance of large logarithms, thus, under excellent theoretical control. Having established the stability of the NLO QCD results with respect to the $p_T(b)$, $p_T(j)$ and $p_T^{miss}$ cut for the integrated fiducial cross section  we move on to differential fiducial cross section distributions.

% =============================================
%
\section{Differential fiducial cross section distributions}
\label{sec:differential}
%
% =============================================
%
%=============================================
\begin{figure}[t!]
  \begin{center}
    \includegraphics[width=0.49\textwidth]{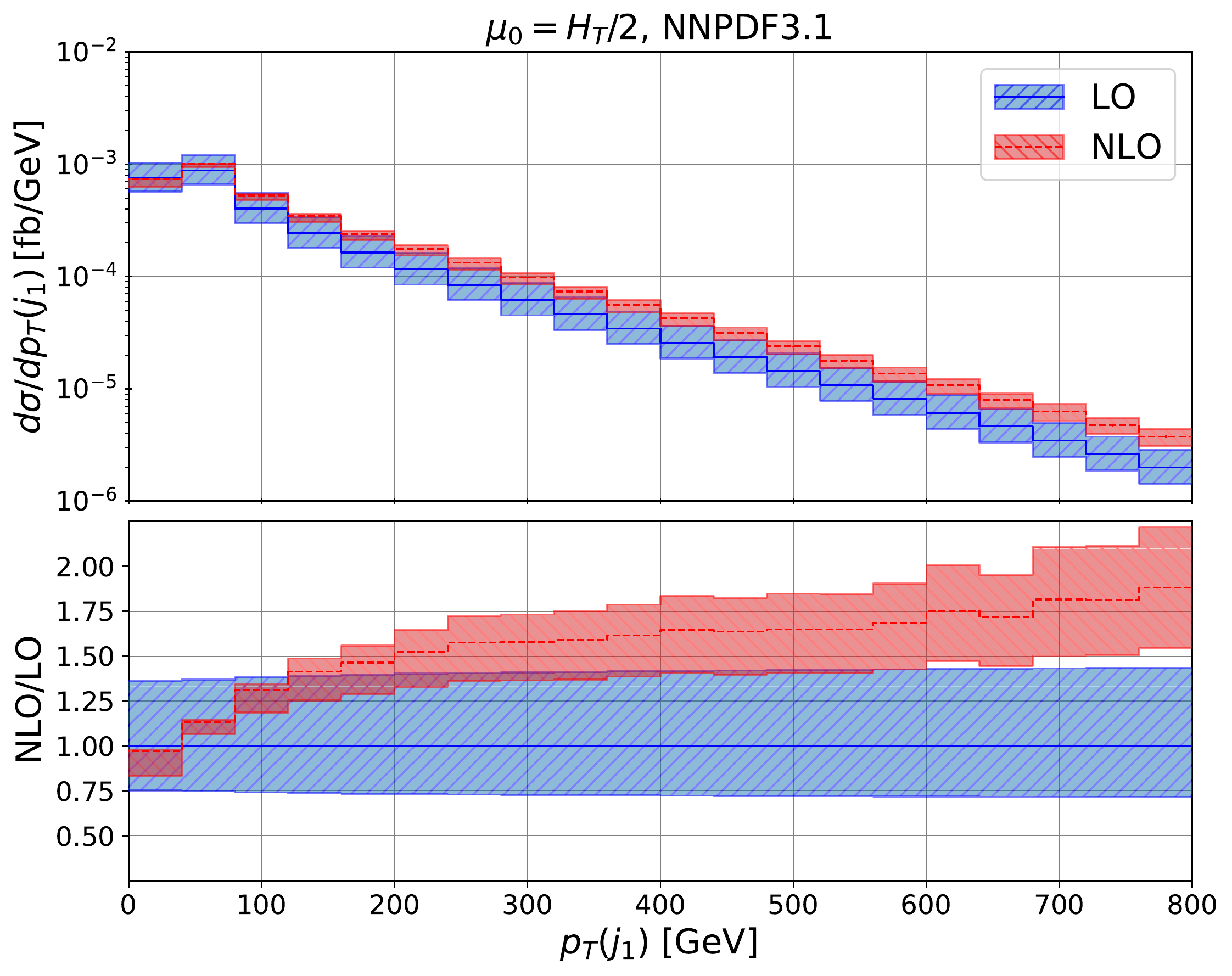}
    \includegraphics[width=0.49\textwidth]{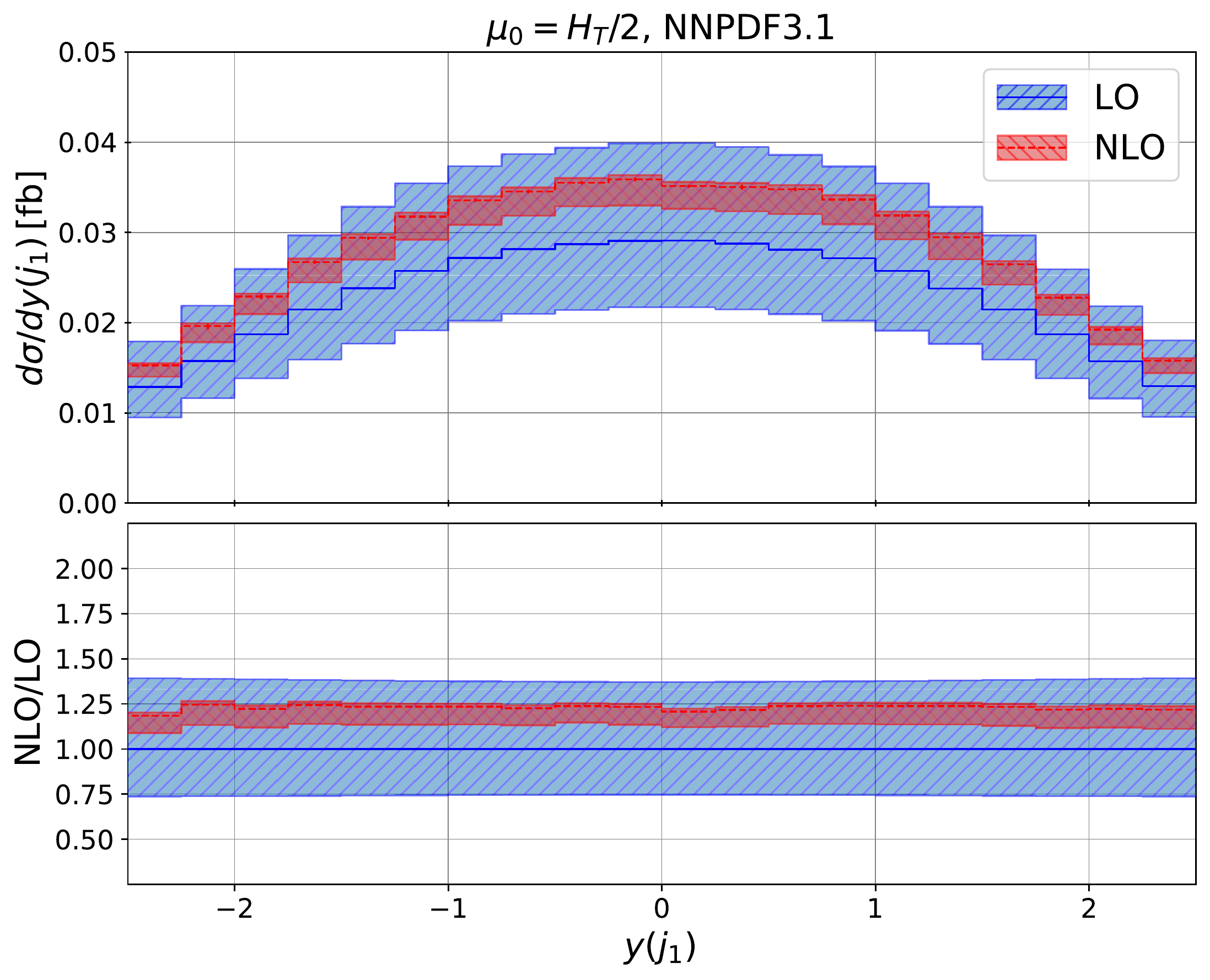}
\end{center}
\caption{\label{fig:Kfactor_ptj_yj} \it 
Differential cross section distributions at LO and NLO in QCD as a function of $p_T(j_1)$ and  $y(j_1)$ for the  $pp \to \WWW \, j+X$ process at the LHC with $\sqrt{s}= 13$ TeV. The upper plots show absolute predictions together with their corresponding uncertainty bands. The lower panels display the differential ${\cal K}$-factor together with its uncertainty band and the relative scale uncertainties of the LO cross section. Results are provided for $\mu_0=H_T/2$ and  (N)LO NNPDF3.1 PDFs.}
 \end{figure}
%=============================================
\begin{figure}[t]
  \begin{center}
    \includegraphics[width=0.49\textwidth]{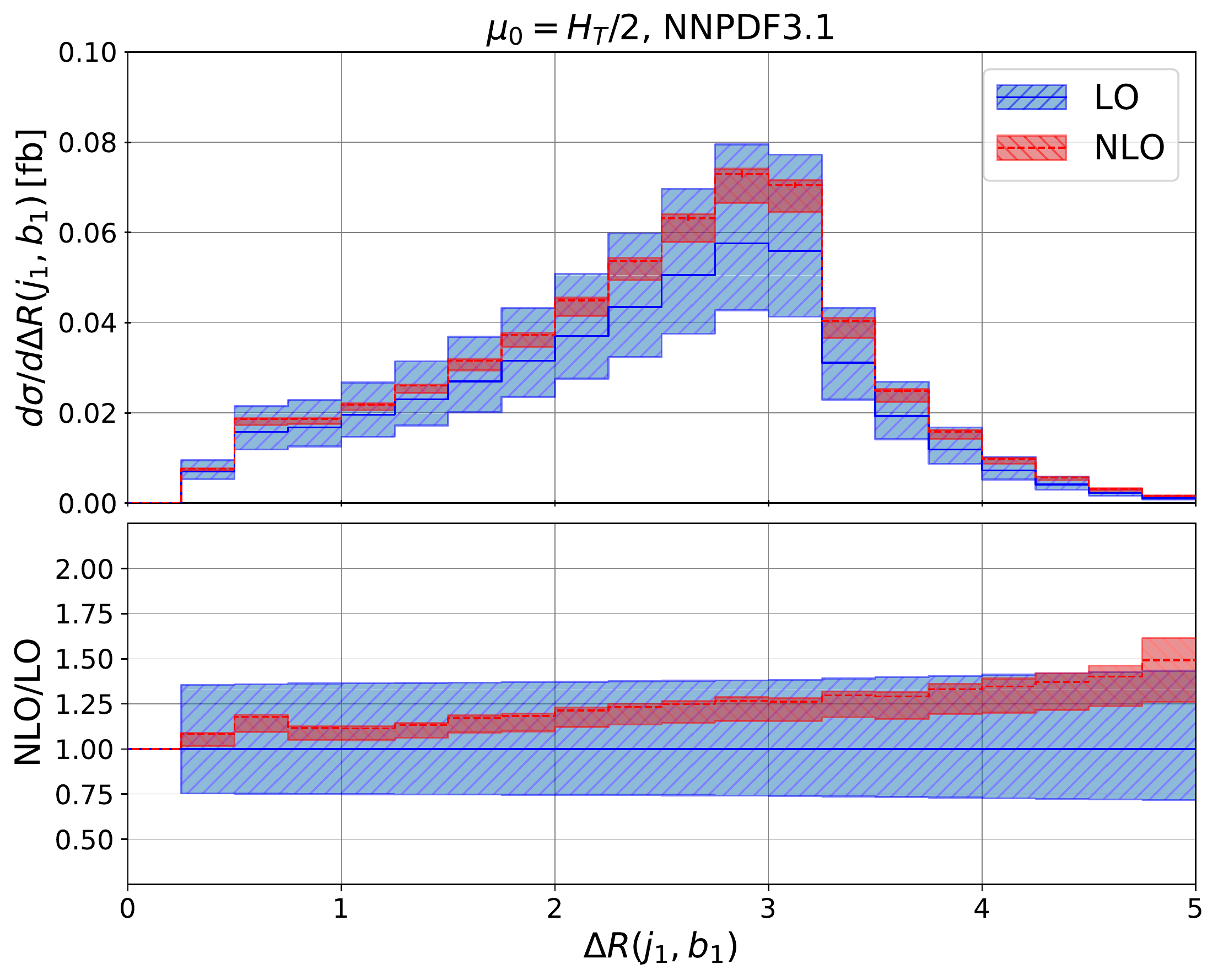}
    \includegraphics[width=0.49\textwidth]{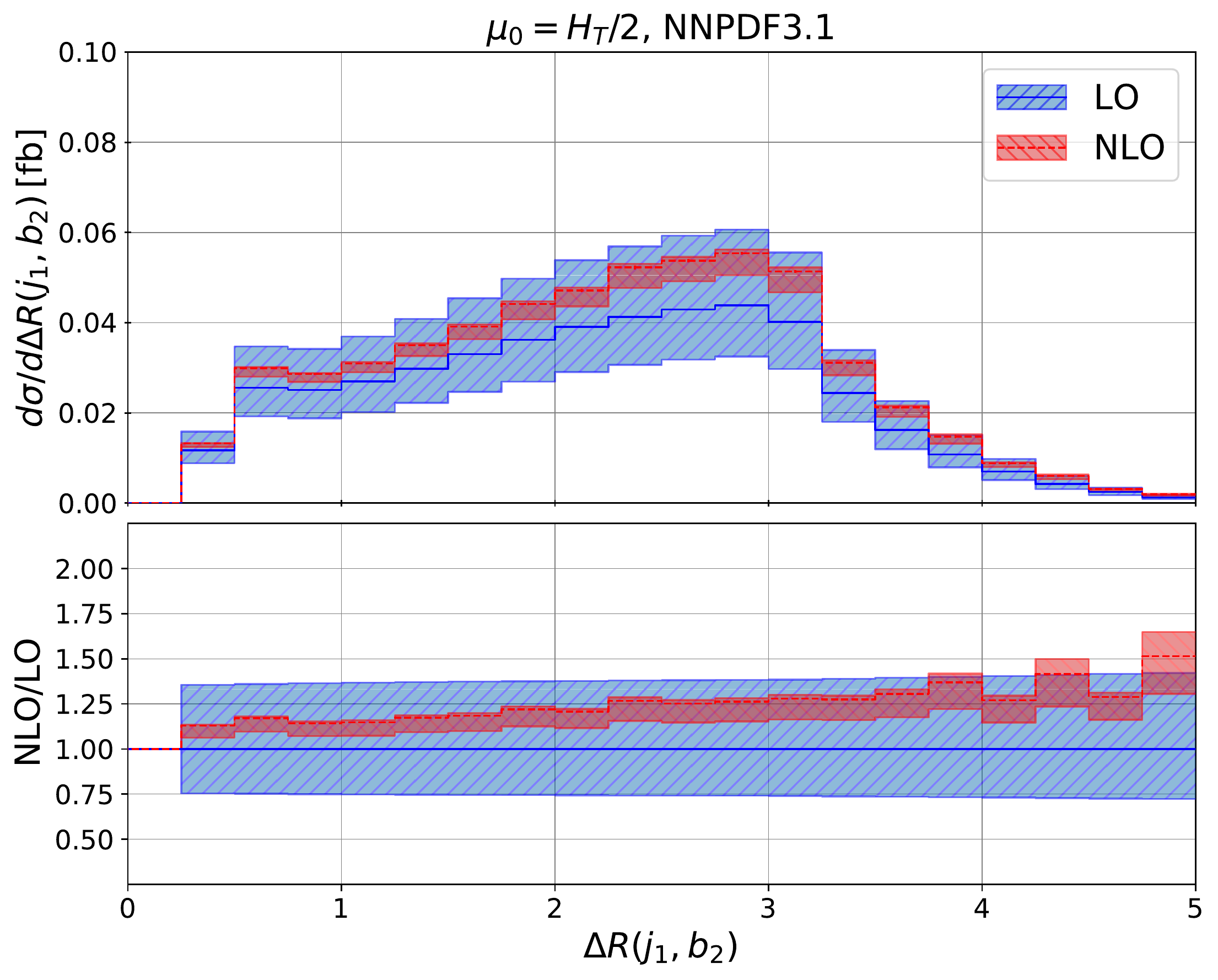}
\end{center}
\caption{\label{fig:Kfactor_drjb} \it Differential cross section
  distributions at LO and NLO in QCD 
  as a function of $\Delta R(j_1,b_1)$ and $\Delta R(j_1,b_2)$
  for the  $pp \to \WWW \, j+X$ process at the LHC with $\sqrt{s}= 13$ TeV.
The upper plots show absolute predictions together with their corresponding uncertainty bands. The lower panels display the differential ${\cal K}$-factor together with its uncertainty band and the relative scale uncertainties of the LO cross section. Results are provided for  $\mu_0=H_T/2$ and  (N)LO NNPDF3.1 PDFs. }
 \end{figure}
 % =============================================
 \begin{figure}[t]
  \begin{center}
    \includegraphics[width=0.49\textwidth]{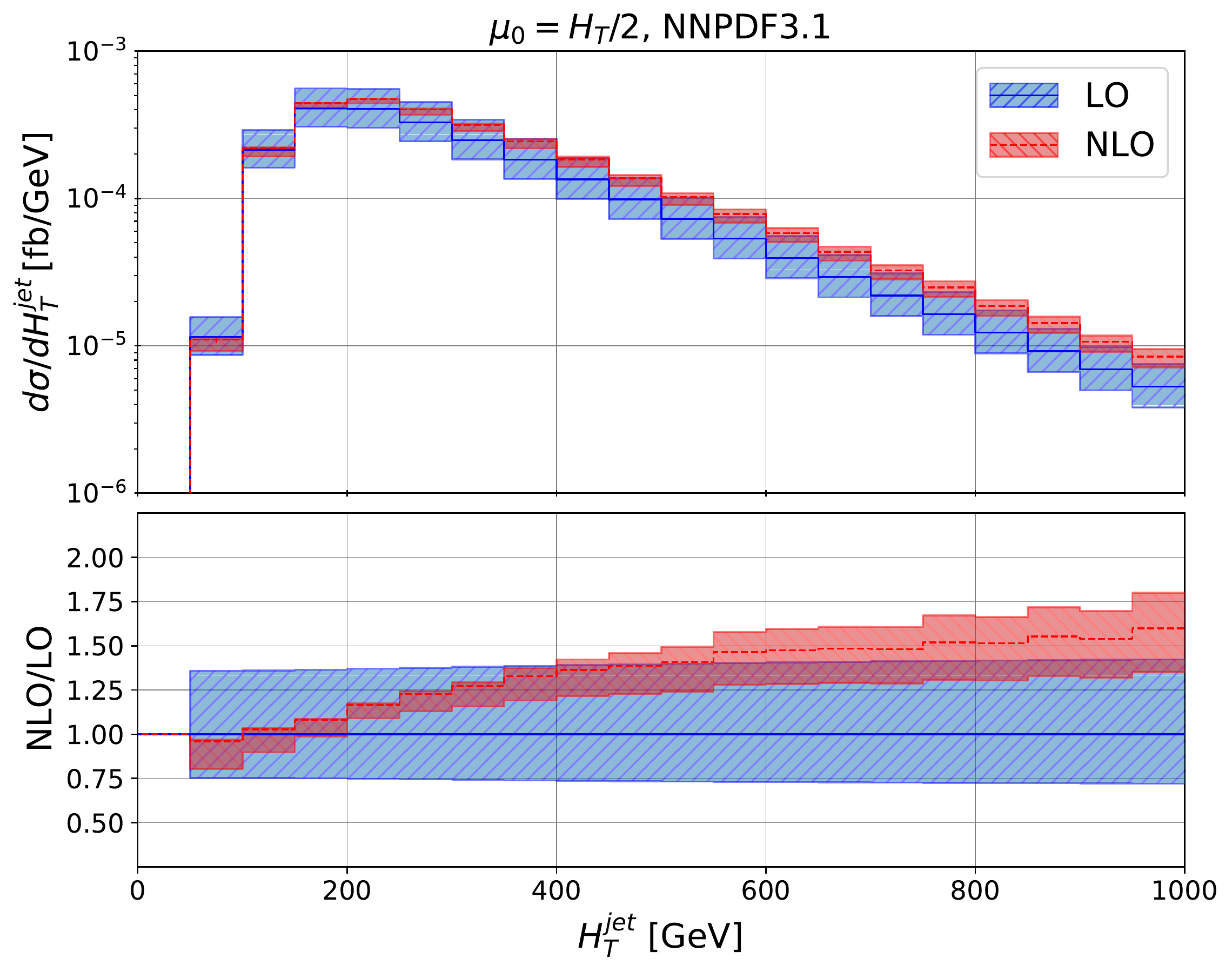}
    \includegraphics[width=0.49\textwidth]{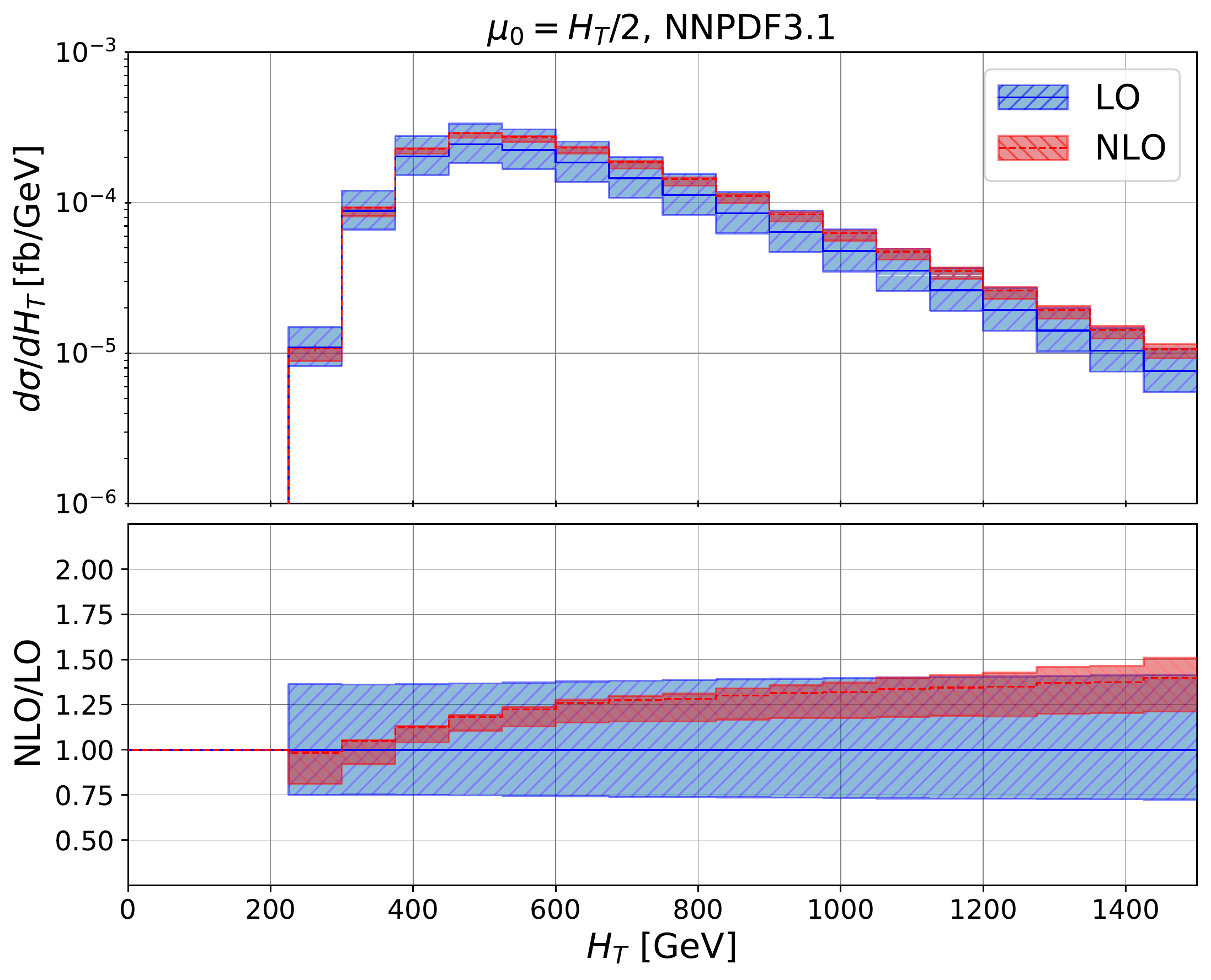}
\end{center}
\caption{\label{fig:Kfactor_htjet_ht} \it 
Differential cross section
  distributions at LO and NLO in QCD as a function of $H_T^{jet}$ and $H_T$
  for the  $pp \to \WWW \, j+X$ process at the LHC with $\sqrt{s}= 13$ TeV.
The upper plots show absolute  predictions together with their corresponding uncertainty bands. The lower panels display the differential ${\cal K}$-factor together with its uncertainty band and the relative scale uncertainties of the LO cross section. Results are provided for  $\mu_0=H_T/2$ and (N)LO NNPDF3.1 PDFs. }
 \end{figure}
%=============================================

The impact of NLO QCD corrections can be significantly enhanced
in specific regions of the phase space compared to the full fiducial region of
the phase space. This can be observed in the shape variation of several differential cross section distributions. To examine this, we have studied about 80 differential cross section distributions and collected the six most interesting results in this Section. Furthermore,  we are going to discuss theoretical  uncertainties associated with these observables. We present our findings for the default setup with  $\mu_R=\mu_F=\mu_0=H_T/2$ as well as with LO and NLO NNPDF3.1 PDF sets. In each case upper panels display absolute LO (blue) and NLO (red) predictions together with corresponding uncertainty bands resulting from 7-point scale variations. The latter is assessed on a bin-by-bin basis. In lower panels we show the same LO and NLO predictions normalized to LO results at $\mu_0$. Thus, the blue band illustrates the relative scale uncertainty of the LO cross section, while the central curve of the red band corresponds to the usual  differential  ${\cal K}$-factor.  In Figure \ref{fig:Kfactor_ptj_yj} we show 
LO and NLO differential cross section distributions as a function of the transverse momentum and rapidity of the light jet, denoted as $p_T(j_1)$ and $y(j_1)$ respectively. In the case of two resolved light jets that pass all the cuts  the hardest one in $p_T$ is plotted. For the dimensionful observable large NLO QCD corrections are obtained up to even $80\%-90\%$ in the tail. This shows the importance of higher-order corrections in describing additional light jet activity in the $pp\to t\bar{t}W^+ \, j$ process and in the $pp\to t\bar{t}W^+$ one for that matter. In the latter case this additional light jet if present would be modeled at best only at LO. The NLO scale uncertainties for $p_T(j_1)$ are below $18\%$ within the  whole plotted range. This should be compared to the LO ones that are consistently at the $40\%$ level. In Figure \ref{fig:Kfactor_ptj_yj} we also present the rapidity of the hardest light jet. Unlike the $b$-jets, the light jet does not show a preference for the central region of the detector and is more uniformly distributed in the $y(j_1)\in (-2.5,2.5)$ range. Furthermore, in this case  an almost constant differential ${\cal K}$-factor is obtained. In details, NLO QCD corrections are of the order of $20\%-25\%$  while the corresponding uncertainties are up to $10\%$ only. Thus, a reduction of the theoretical error by a factor of $4$ is observed for this dimensionless observable when higher-order effects are incorporated. Unlike the $p_T(j_1)$ distribution for $y(j_1)$ the NLO result is well within the LO uncertainty bands. Thus, in this case the proposed dynamical scale can be used within a LO calculation, together with a suitably chosen global ${\cal K}$-factor, to obtain results that well approximate the full NLO QCD result for $y(j_1)$. However, we will see that this is not the case for other dimensionless differential cross section  distributions that can be constructed for the light jet.  

In Figure \ref{fig:Kfactor_drjb} we display the angular separation in the azimuthal-angle-rapidity plane between the light jet and the hardest $b$-jet as well as between the light jet and the second hardest $b$-jet denoted as  $\Delta R(j_1,b_1)$ and $\Delta R(j_1,b_2)$ respectively. We can observe that the light jet and the hardest  $b$-jet are predominantly produced in the back-to-back configuration. We have checked using the reconstruction method 
described in Section  \ref{sec:inputparameters} that about $90\%$ of light jets are emitted in the $t\bar{t}Wj$ production stage, which is in line
with our observation here. For the light jet and the second hardest $b$-jet the spectrum  is smeared out and the enhancement towards  smaller values of  $\Delta R(j_1,b_2)$ can also be noticed. For $\Delta R(j_1,b_1)$ up to about $\Delta R(j_1,b_1) \sim 3$ higher-order corrections are in the range of  $10\%-30\%$. They increase, however, up to $40\%-50\%$ in the vicinity of  $\Delta R(j_1,b_1) \sim 5$. Also the size of theoretical uncertainties due to the scale dependence is rather different in these two regions. For   $\Delta R(j_1,b_1)\lesssim 3$ they  are below $8\%$ and  rise to  $15\%$ for $\Delta R(j_1,b_1)\sim 5$. Qualitatively similar results are obtained for the  $\Delta R(j_1,b_2)$ observable. 

Lastly, in Figure  \ref{fig:Kfactor_htjet_ht} we provide our findings for the scalar sum of the transverse momenta of all jets in the process denoted as $H_T^{jet}$ and given by
\begin{equation}
H_T^{jet} = p_T(b_1)+ p_T(b_2) + p_T(j_1)\,. 
\end{equation}
Also shown is the $H_T$ observable already defined in Eq. \eqref{eq:ht}.  
In both cases, the LO uncertainties are at a rather constant level of $40\%$ and are significantly reduced when the NLO QCD corrections are calculated. In particular, they can reach a maximum of $16\%-18\%$. In addition, for both observables, higher-order effects are substantial. NLO QCD corrections can reach $60\%$ in the tail for $H_T^{jet}$, whereas  for the $H_T$ observable, which in addition comprises the three charged leptons and the missing transverse momentum, they are decreased to $40\%$.

% =============================================
 \begin{figure}[t]
  \begin{center}
    \includegraphics[width=0.49\textwidth]{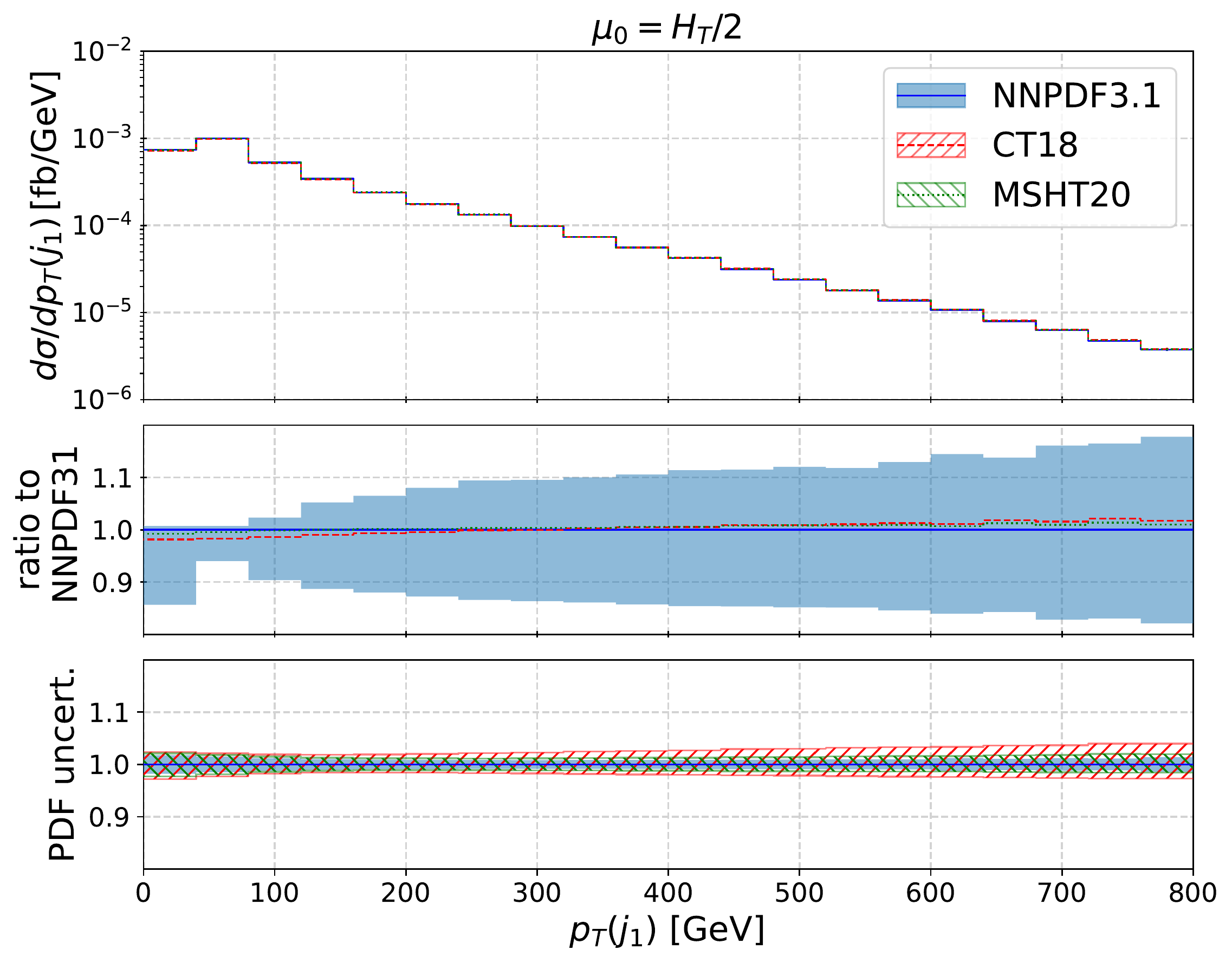}
    \includegraphics[width=0.49\textwidth]{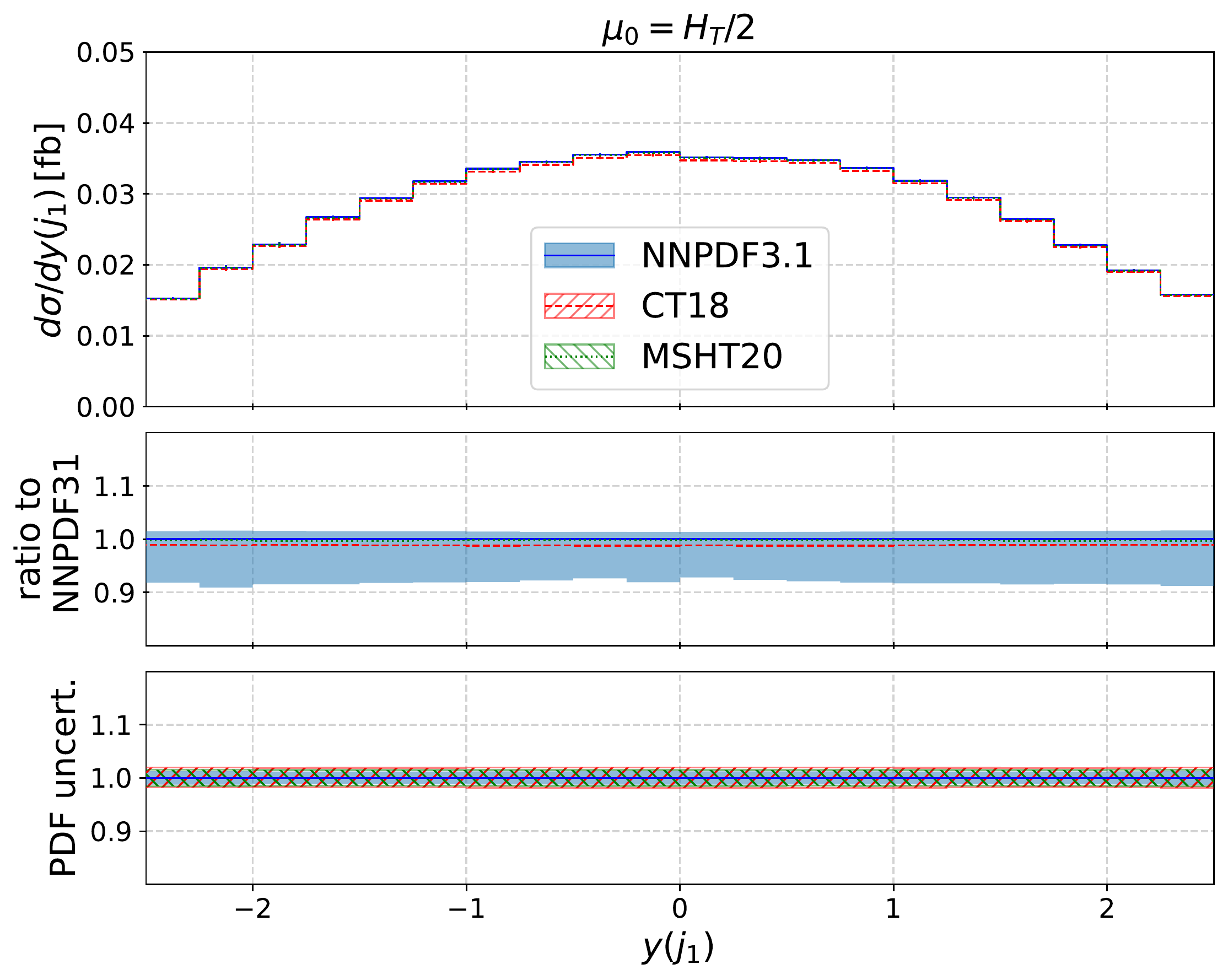}
\end{center}
\caption{\label{fig:PDF_ptj_yj} \it  Differential cross section
  distributions at  NLO in QCD as a function of $p_T(j_1)$ and $y(j_1)$ for the  $pp \to \WWW \, j+X$ process at the LHC with $\sqrt{s}= 13$ TeV. The upper panel shows the absolute predictions for the following  NLO PDF sets: NNPDF3.1,  CT18  and MSHT20. Results are given for $\mu_R = \mu_F = \mu_0 = H_T/2$. The middle panel displays the ratio to the result with the default NNPDF3.1 PDF set as well as its scale dependence. The lower panel presents the internal PDF uncertainties calculated separately for each PDF set.}
 \end{figure}
 % =============================================
  \begin{figure}[t]
  \begin{center}
    \includegraphics[width=0.49\textwidth]{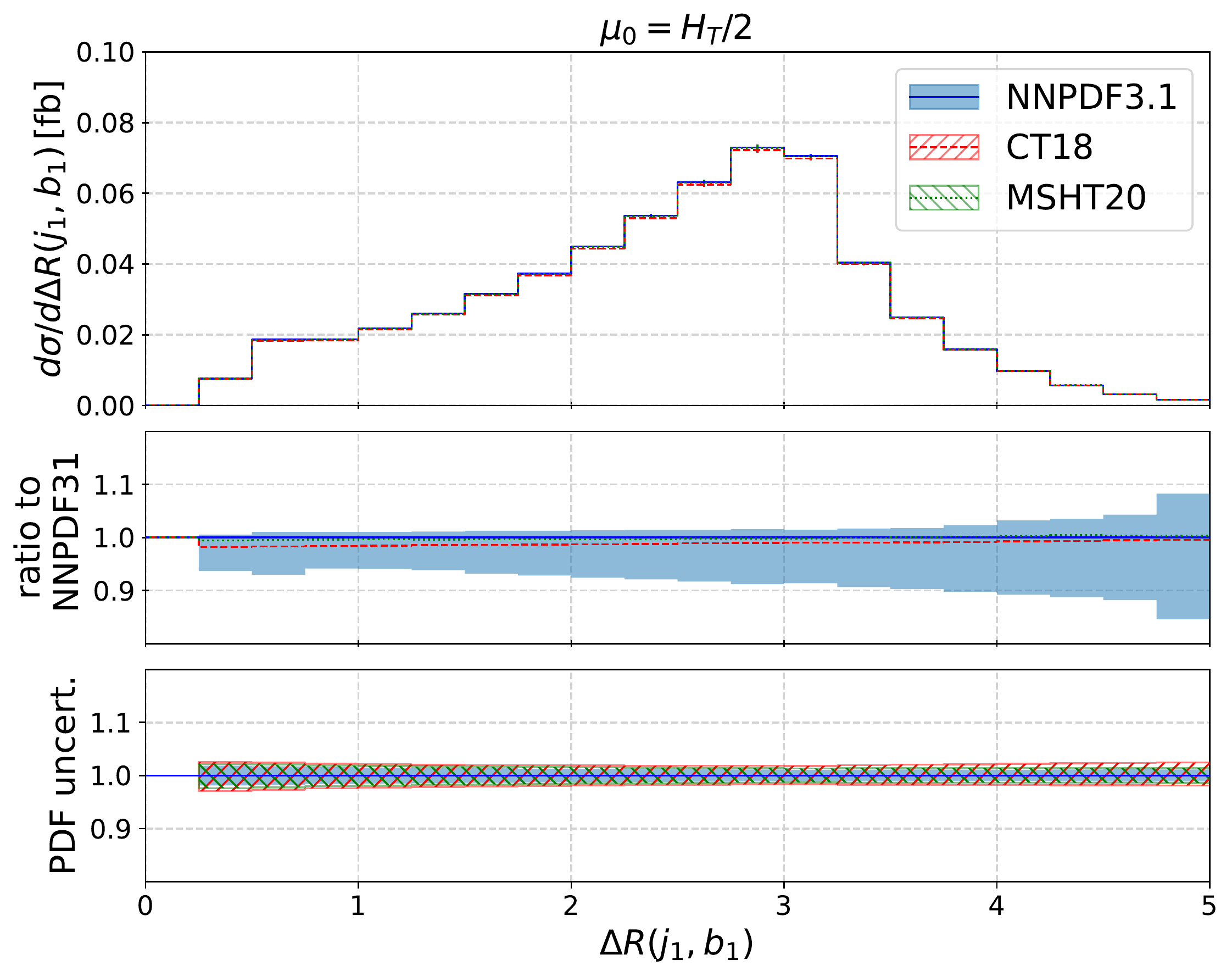}
    \includegraphics[width=0.49\textwidth]{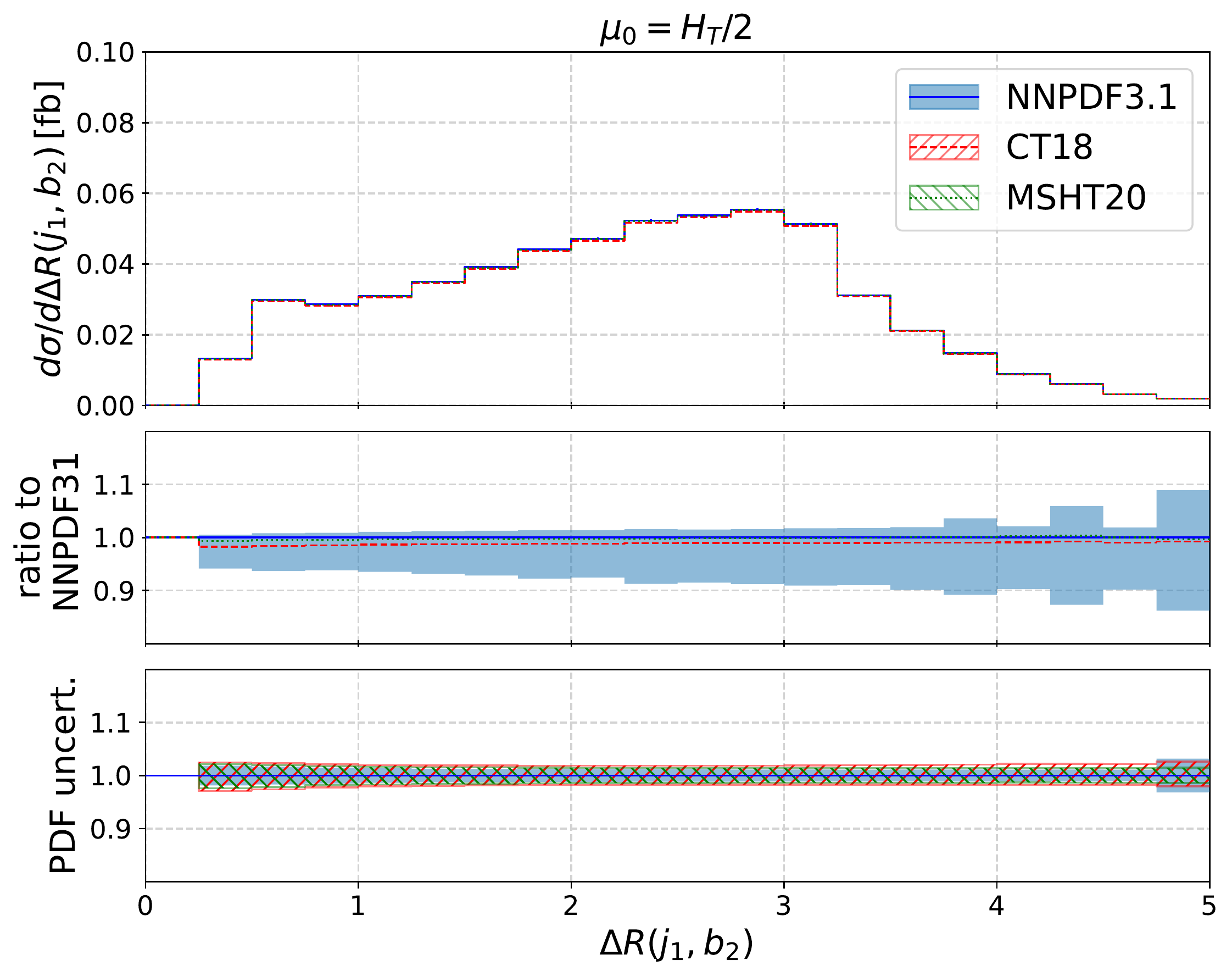}
\end{center}
\caption{\label{fig:PDF_drjb} \it  Differential cross section
  distributions at  NLO in QCD as a function of $\Delta R(j_1,b_1)$ and $\Delta R(j_1,b_2)$ for the  $pp \to \WWW \, j+X$ process at the LHC with $\sqrt{s}= 13$ TeV. The upper panel shows the absolute predictions for the following NLO PDF sets: NNPDF3.1,  CT18  and MSHT20. Results are given for $\mu_R = \mu_F = \mu_0 = H_T/2$. The middle panel displays the ratio to the result with the default NNPDF3.1 PDF set as well as its scale dependence. The lower panel presents the internal PDF uncertainties calculated separately for each PDF set.}
 \end{figure}
 % =============================================
  \begin{figure}[t]
  \begin{center}
    \includegraphics[width=0.49\textwidth]{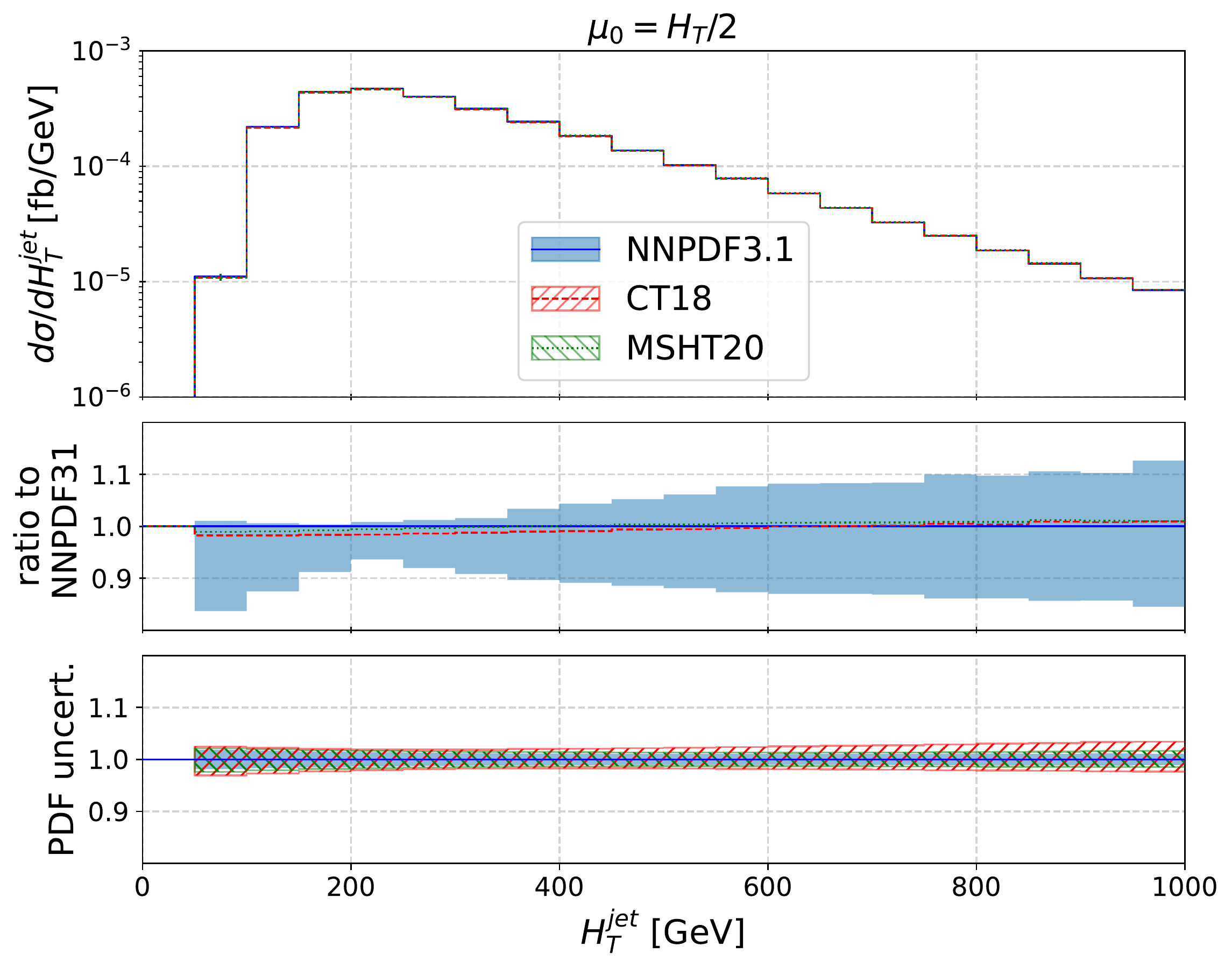}
    \includegraphics[width=0.49\textwidth]{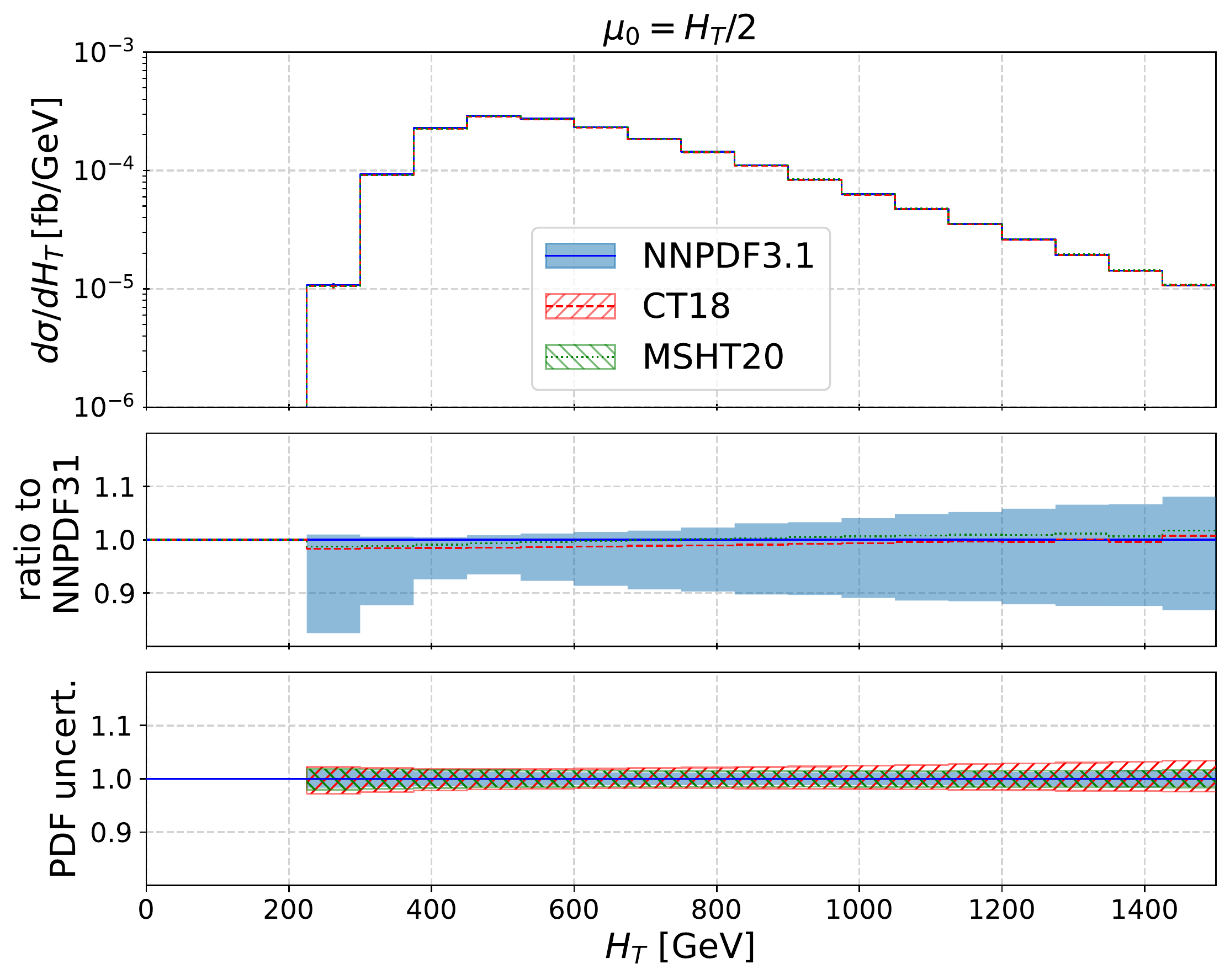}
\end{center}
\caption{\label{fig:PDF_htjet_ht} \it  Differential cross section
  distributions at  NLO in QCD as a function of $H_T^{jet}$ and $H_T$ for the  $pp \to \WWW \, j+X$ process at the LHC with $\sqrt{s}= 13$ TeV. The upper panel shows the absolute predictions for the following NLO PDF sets: NNPDF3.1,  CT18  and MSHT20. Results are given for $\mu_R = \mu_F = \mu_0 = H_T/2$. The middle panel displays the ratio to the result with the default NNPDF3.1 PDF set as well as its scale dependence. The lower panel presents the internal PDF uncertainties calculated separately for each PDF set.}
 \end{figure}
%=============================================

Having examined the magnitude of the theoretical uncertainties due to scale variation for the differential cross section  distributions we turn our attention to the PDF uncertainties. We have already verified that the latter are smaller  at the integrated fiducial cross section  level. We would like to check whether this is also the case at the differential cross section  level. In Figure \ref{fig:PDF_ptj_yj} - Figure \ref{fig:PDF_htjet_ht} we present the same six observables that we have previously studied, but this time we focus on NLO QCD predictions for three different PDF sets: NNPDF3.1, CT18 and MSHT20. All the plots consist of three panels. Upper panels show absolute NLO QCD predictions  for NNPDF3.1, CT18 and MSHT20. Middle panels present the ratios of the three predictions to the result obtained with our default NNPDF3.1 PDF set. Thus, this part highlights the relative differences due to the choice of the PDF set.  The scale uncertainties for the NLO QCD result also with the default NNPDF3.1 PDF set  are additionally given there for comparison. Finally, bottom panels display the internal PDF uncertainty of the three given PDF sets. The internal PDF uncertainty bands are normalized to the central value of the corresponding PDF set. Our NLO QCD results are provided for $\mu_R=\mu_F=\mu_0=H_T/2$. 

To illustrate our findings we will concentrate only on $p_T (j_1)$ and $y(j_1)$ differential cross section  distributions that are displayed in Figure 
\ref{fig:PDF_ptj_yj} as for other observables    similar conclusions can be reached. For the $p_T (j_1)$ observable the relative differences between MSHT20 and CT18 relative to NNPDF3.1  are maximally up to $1.5\%$ for MSHT20 and  $2.2\%$ for CT18. The internal PDF uncertainties for this differential cross section distributions are the largest in the tail and are of the order of $1.8\%$, $4\%$ and $2.3\%$ for NNPDF3.1, CT18 and MSHT20 respectively. Furthermore, we note that, with the exception of the CT18 PDF set, the relative PDF uncertainties are of similar magnitude as the internal PDF uncertainties obtained for NNPDF3.1 and MSHT20. This should be contrasted with the scale uncertainties, which are at the $18\%$ level in these fiducial phase-space regions. Thus, the PDF uncertainties remain negligible compared to the NLO scale variation. This observation is also true for dimensionless observables. Indeed, for $y(j_1)$, for which the theoretical uncertainties due to scale variation are maximally up to $10\%$, we note even more consistent results for the three PDF sets. The PDF uncertainties are up to $1.3\%$, $2\%$ and $1.6\%$ correspondingly for NNPDF3.1, CT18 and MSHT20.  The relative differences among these three PDF sets are below $1.2\%$.  We conclude this section by stating that, also at the differential cross section level, the PDF uncertainties are very small compared to the corresponding theoretical uncertainties arising from the scale dependence. 

% =============================================
%
\section{Additional jet activity in $\boldsymbol{t\bar{t} W^+}$ production}
\label{sec:comptottw}
%
% =============================================

To investigate the importance of additional jet activity in the associated production of a top-quark pair with a $W$ gauge boson at the LHC we compare the $pp \to \WWW \,j +X$ process with the simpler one $pp \to \WWW \, +X$. Also in the latter case full off-shell top-quark and $W$ gauge boson effects are incorporated at the matrix element level. We note here, that in Refs. \cite{Bevilacqua:2020pzy,Bevilacqua:2020srb,Bevilacqua:2021tzp} the $pp \to t\bar{t}W^+$ production process in the multi-lepton channel with full off-shell  effects has  been studied, however, the following final state $pp \to e^+\nu_e\, \mu^- \bar{\nu}_\mu \,e^+ \nu_e \, b\bar{b} +X$ has been investigated there instead.  Thus, our predictions for $pp \to \WWW \, +X$ are generated afresh for the setup specified in Section \ref{sec:inputparameters}. The renormalisation and factorisation scale setting, however, are slightly different. Because for the $pp \to \WWW \, +X$  process the leading light jet is only  present in the real-emission part of the calculation we use the following definitions for $H_T$ and $E_T$ instead 
\begin{equation}
\begin{split}
H_T &= p_T(e^+ )+ p_T(\tau^+)+ p_T(\mu^-) +p_{T}^{miss} + p_{T} (b_1) +
p_{T} (b_2) \,,\\[0.2cm]
E_T &= \sqrt{m_t^2 + p_{T}^2 (t)} + \sqrt{m_t^2 + p_{T}^2(\tb \,)} +
\sqrt{m_W^2 + p_{T}^2 (W)}  \,.
\end{split}
\end{equation}
In short, the transverse momentum of the light jet, $p_T(j_1)$, is omitted in both definitions for the  $pp \to \WWW \, +X$ process. In line with our previous work we employ $\mu_R=\mu_F=\mu_0=H_T/3$ and $\mu_R=\mu_F=\mu_0=E_T/3$. 
%
%=============================================
\begingroup
\begin{table}[t]
    \centering
    \begin{tabular}{l l l l l l l}
        \hline\hline
        &&&&&&\\[-0.2cm]
         & $\sigma^{t\tb W^+}_{H_T/3}\,[\textrm{ab}]$ & $\sigma^{t\tb W^+j}_{H_T/2}\,[\textrm{ab}]$ & $\sigma^{t\tb W^+}_{E_T/3}\,[\textrm{ab}]$ & $\sigma^{t\tb W^+j}_{E_T/2}\,[\textrm{ab}]$ & $\sigma^{t\tb W^+}_{m_t+m_W/2}\,[\textrm{ab}]$ & $\sigma^{t\tb W^+j}_{m_t+m_W/2}\,[\textrm{ab}]$\\
         &&&&&&\\[-0.2cm]
         \hline\hline
         &&&&&&\\[-0.2cm]
     LO  & $ 216.6^{\,+24\%}_{\,-18\%} $ 
     & $ 115.8^{\,+38\%}_{\,-26\%} $ 
     & $ 198.7^{\, +23 \% }_{ \,-18 \% }$ 
     & $103.9^{\, +37 \% }_{ \,-25 \% }$ 
     & $ 202.6^{ \,+24 \% }_{\, -18 \% }$ 
     & $141.0^{ \,+41 \% }_{ \,-27 \% }$\\[0.2cm]
     NLO & $254.6^{ \,+2.8 \% }_{ \,-5.9 \% }$ 
     & $ 142.3^{\, +1.4 \% }_{ \,-8.1 \% }$ 
     & $ 249.6^{ \,+4.6 \% }_{\, -6.8 \% }$ 
     & $ 139.7^{ \,+3.7 \% }_{\, -9.9 \% }$ 
     & $ 252.3^{\, +4.5 \% }_{ \,-6.8 \% }$ 
     & $144.3^{\, +~0.3 \% }_{ \,-14.1 \% }$\\
     &&&&&&\\[-0.2cm]
     \hline\hline
    \end{tabular}
    \caption{\it Integrated fiducial
      cross sections at LO and NLO in QCD for the $pp \to \WWW \, +X$   and $pp \to \WWW \, j+X$$ $ process  at the LHC with $\sqrt{s}= 13$ TeV.  For the sake of brevity we refer to these processes as  $t\bar{t}W^+$ and $t\bar{t}W^+j$ respectively. Results are provided for three scale settings and for   (N)LO NNPDF3.1 PDFs. The theoretical uncertainties coming from the 7-point scale variation are also shown.}
    \label{tab:ttwttwjfiducial}
\end{table}
\endgroup
%=============================================

Our findings for LO and NLO integrated fiducial cross sections for 
the $pp \to \WWW+X$  and $pp \to \WWW\, j+X$ process are shown in Table \ref{tab:ttwttwjfiducial}.  In each case theoretical predictions are given for three scale settings and for the (N)LO NNPDF3.1 PDF set. In addition, the theoretical uncertainties arising from scale variation are also provided. We can notice that the cross section for the $pp \to \WWW\, j+X$ process  is substantial  compared to the one for  $pp \to \WWW +X$. Indeed, the size of the $pp\to \WWW\, j$  contribution in the inclusive $pp \to \WWW $ sample is  at the level of  $50\%-70\%$  depending on the scale choice and the order in the perturbative expansion in $\alpha_s$. From an experimental point of view, additional jets produced in association with the $\WWW$ system need to be understood very precisely since their appearance affects the reconstruction of the $t\bar{t}W^+$ event.  This additional jet activity can be used not only to examine the underlying production and decay mechanisms even further but also to design new methods for a sizeable reduction of QCD backgrounds. This, of course, increases the importance of the  current study for the $pp \to \WWW\, j+X$ process and the need  to describe this multi-particle  final state as precisely as possible. In addition, in the absence of a result with NNLO QCD corrections for the $pp\to t\bar{t}W^+$ production process a  simulation of NLO $\WWW  + \WWW\, j$  merged sample using for example the \textsc{MiNLO} method \cite{Hamilton:2012np} should be more appropriate  to provide reliable predictions for the underlining process $pp \to \WWW+X$ alone.  We plan to carry out such a study including  top-quark full off-shell effects in the near future. Further scrutinising  the results presented in Table \ref{tab:ttwttwjfiducial} we observe an increase in NLO theoretical uncertainties  from $6\%-7\%$ for $pp\to \WWW+X$ to $8\%-14\%$ in the case of  $pp \to \WWW\, j+X$ depending on the choice of $\mu_0$.  For both dynamical scale settings also  the ${\cal K}$-factor increases when an additional light jet is added to the process at NLO in QCD. Indeed, for  the $H_T \,(E_T)$ based scales we observe a change from ${\cal K}= 1.18$ to ${\cal K}= 1.23$ (from ${\cal K}= 1.26$ to ${\cal K}= 1.34$). For our fixed scale choice  NLO QCD corrections are actually substantially reduced for $pp \to \WWW\, j+X$ and are of the order of $2\%$ only comparing to $25\%$ for $pp \to \WWW+X$. This effect is driven by the larger LO cross section for the $pp \to \WWW \, j$ process as compared to the LO result for this process either with $\mu_0=H_T/2$ or $\mu_0=E_T/2$. 
%
%=============================================
\begingroup
\begin{table}[t]
    \centering
    \begin{tabular}{c c c c}
    \hline \hline
    &&&\\[-0.2cm]
    $(\mu_R, \mu_F)$ & $\sigma^{t\bar{t}W^+}_{H_T/3} (N_j=0)$ [ab] &
          $\sigma^{t\bar{t}W^+}_{H_T/3} (N_j=1)$ [ab] &
          $\sigma^{t\bar{t}W^+}_{H_T/3} (N_j \ge 0)$ [ab] \\
          &&&\\[-0.2cm]
        \hline\hline
        &&&\\[-0.2cm]
($\mu_0,\mu_0$)      &  $81.8{}^{\, +41\%}_{\, -84\%}$ 
&  $172.8{}^{\, +44\%}_{\, -28\%}$ 
&  $254.6{}^{\, +2.8\%}_{\, -5.9\%}$\\
&&&\\[-0.2cm]
\hline \hline
&&&\\[-0.2cm]
$(\mu_R, \mu_F)$ & $\delta \sigma^{t\bar{t}W^+}_{H_T/3} (N_j=0)$ [ab] &
          $\delta \sigma^{t\bar{t}W^+}_{H_T/3} (N_j=1)$ [ab] &
          $ \delta \sigma^{t\bar{t}W^+}_{H_T/3} (N_j\ge 0)$ [ab]\\
          &&&\\[-0.2cm]
          \hline\hline 
          &&&\\[-0.2cm]
($2\mu_0,2\mu_0$)    & $+33.8 \, (+41 \%)$ 
& $-48.7 \,(-28 \%)$ & $-14.9 \,(-5.9 \%)$ \\[0.2cm]
($\mu_0/2,\mu_0/2$)  & $-68.5 \,(-84 \%)$ 
& $+75.3 \, (+44 \%)$ 
& $ \, +6.8 \, (+2.7 \%)$\\[0.2cm]
($2\mu_0,\mu_0$)     & $+31.1 \, (+38 \%)$ 
& $-41.2 \, (-24 \%)$ 
& $-10.1\, (-4.0 \%)$\\[0.2cm]
($\mu_0/2,\mu_0$)    & $-61.8 \,(-76 \%)$ 
& $+60.8 \,(+35 \%)$ & $\, -1.0\, (-0.4 \%)$\\[0.2cm]
($\mu_0,2\mu_0$)     & $\, +5.0 \,(+6.1 \%)$ 
& $-9.8\, (-5.7 \%)$ & $\, -4.7 \,(-1.8 \%)$\\[0.2cm]
($\mu_0,\mu_0/2$)    & $\, -3.6\, (-4.4 \%)$ 
& $+10.8 \,(+6.3 \%)$ & $\, +7.2 \,(+2.8 \%)$\\
&&&\\[-0.2cm]
\hline \hline
    \end{tabular}
       \caption{\it 
         Integrated fiducial cross sections at NLO in QCD as a function of the number of resolved
         light jets (denoted as $N_j$) for the $pp \to \WWW \, +X$  process at the LHC with
         $\sqrt{s}= 13$ TeV.  For the sake of brevity we refer to this processes as  $t\bar{t}W^+$.
         Results are provided for $\mu_0=H_T/3$ and the NLO NNPDF3.1 PDF set. The theoretical
         uncertainties coming from the 7-point scale variation are also shown.}
    \label{tab:ttwNjet}
\end{table}
\endgroup
%=============================================
\begingroup
\begin{table}[t]
    \centering
    \begin{tabular}{c c c c}
\hline \hline
&&&\\[-0.2cm]
    $(\mu_R, \mu_F)$ & $\sigma^{t\bar{t}W^+j}_{H_T/2} (N_j=1)$ [ab] &
          $\sigma^{t\bar{t}W^+ j}_{H_T/2} (N_j=2)$ [ab] &
          $\sigma^{t\bar{t}W^+ j}_{H_T/2} (N_j\ge 1)$ [ab]\\
          &&&\\[-0.2cm]
\hline\hline
&&&\\[-0.2cm]
($\mu_0,\mu_0$)      &  $78.6{}^{\,+13\% }_{\,-48\%}$ 
& $63.7{}^{\,+56\%}_{\,-34\%}$ 
&  $142.3{}^{\,+1.4\% }_{\, -8.1\%}$ \\
&&&\\[-0.2cm]
\hline\hline
&&&\\[-0.2cm]
$(\mu_R, \mu_F)$ & $ \delta \sigma^{t\bar{t}W^+j}_{H_T/2} (N_j=1)$ [ab] &
          $\delta \sigma^{t\bar{t}W^+ j}_{H_T/2} (N_j=2)$ [ab] &
          $ \delta \sigma^{t\bar{t}W^+ j}_{H_T/2} (N_j\ge 1)$ [ab]\\
          &&&\\[-0.2cm]
          \hline\hline 
          &&&\\[-0.2cm]
($2\mu_0,2\mu_0$)    & $+9.9\, (+13 \%)$ 
& $-21.4 \, (-34 \%)$ & $-11.5 \,(-8.1 \%)$ \\[0.2cm]
($\mu_0/2,\mu_0/2$)  & $-37.4 \,(-48 \%) $
& $+35.7\, (+56 \%)$ & $\, -1.6 \, (-1.1 \%)$ \\[0.2cm]
($2\mu_0,\mu_0$)     & $\, +9.0 \,(+11 \%) $
& $-18.0 \, (-28 \%)$ & $\, -9.0  \,(-6.3 \%)$ \\[0.2cm]
($\mu_0/2,\mu_0$)    & $-28.7 \,(-37 \%)$ 
& $+27.8 \,(+44 \%)$ & $\, -0.9 \,(-0.6 \%)$  \\[0.2cm]
($\mu_0,2\mu_0$)     & $\, +3.5  \,(+4.5 \%)$ 
& $\, -4.8 \, (-7.5 \%)$ & $\, -1.2 \,(-0.8 \%)$ \\[0.2cm]
($\mu_0,\mu_0/2$)    & $\, -3.5 \,(-4.5 \%) $
& $\, +5.5 \, (+8.6 \%)$ & $\, +2.0 \, (+1.4 \%)$  \\
&&&\\[-0.2cm]
     \hline\hline
    \end{tabular}
       \caption{\it 
      Integrated fiducial cross sections at NLO in QCD as a function of the number of resolved light jets (denoted as $N_j$) for the  $pp \to \WWW \, j+X$ process at the LHC with $\sqrt{s}= 13$ TeV. 
      For the sake of brevity we refer to this processes as $t\bar{t}W^+j$. Results are provided for $\mu_0=H_T/2$ and for the NLO NNPDF3.1 PDF set. The theoretical uncertainties coming from the 7-point scale variation are also shown.}
    \label{tab:ttwjNjet}
\end{table}
\endgroup
% =============================================
  \begingroup
\begin{table}[t]
    \centering
    \begin{tabular}{c c c c}
\hline \hline
&&&\\[-0.2cm]
    $(\mu_R, \mu_F)$ & $\sigma^{t\bar{t}W^+}_{H_T/2} (N_j=0)$ [ab] &
          $\sigma^{t\bar{t}W^+ }_{H_T/2} (N_j=1)$ [ab] &
          $\sigma^{t\bar{t}W^+ }_{H_T/2} (N_j\ge 0)$ [ab]\\
          &&&\\[-0.2cm]
\hline\hline
&&&\\[-0.2cm]
($\mu_0,\mu_0$)      &  $104.6{}^{\,+20\% }_{\,-44\%}$ 
& $141.9 {}^{\,+41\%}_{\,-27\%}$ 
&  $246.4{}^{\,+5.0\% }_{\, -7.0\%}$ \\
&&&\\[-0.2cm]
\hline\hline
&&&\\[-0.2cm]
$(\mu_R, \mu_F)$ & $ \delta \sigma^{t\bar{t}W^+}_{H_T/2} (N_j=0)$ [ab] &
          $\delta \sigma^{t\bar{t}W^+ }_{H_T/2} (N_j=1)$ [ab] &
          $ \delta \sigma^{t\bar{t}W^+ }_{H_T/2} (N_j\ge 0)$ [ab]\\
          &&&\\[-0.2cm]
          \hline\hline 
          &&&\\[-0.2cm]

      ($2\mu_0,2\mu_0$)    & $+21.1\, (+20 \%)$
      & $-38.4\, (-27 \%)$ & $-17.2 \,(-7.0 \%)$ \\[0.2cm]
      ($\mu_0/2,\mu_0/2$)  & $-45.7\, (-44 \%)$
      & $+58.1\, (+41 \%)$ & $+12.4\, (+5.0 \%)$\\[0.2cm]    
      ($2\mu_0,\mu_0$)     & $+19.1\, (+18 \%)$
      & $-32.3 \,(-23 \%)$ & $-13.2\, (-5.3 \%)$\\[0.2cm]
      ($\mu_0/2,\mu_0$)    & $-40.0 \,(-38 \%)$
      & $+46.6\, (+33 \%)$ & $+6.6\, (+2.7 \%)$\\[0.2cm]
      ($\mu_0,2\mu_0$)     & $+4.1 \,(+3.9 \%)$
      & $-7.8\, (-5.5 \%)$ & $-3.8\, (-1.5 \%)$\\[0.2cm]
      ($\mu_0,\mu_0/2$)    & $-3.1\, (-2.9 \%)$
      & $+8.7 \,(+6.1 \%)$ & $+5.6\, (+2.3 \%)$\\
      &&&\\[-0.2cm]
     \hline\hline
    \end{tabular}
      \caption{\it 
        Integrated fiducial cross sections at NLO in QCD as a function of the number of resolved
        light jets (denoted as $N_j$) for the $pp \to \WWW \, +X$  process at the LHC with
        $\sqrt{s}= 13$ TeV.  For the sake of brevity we refer to this processes as  $t\bar{t}W^+$.
        Results are provided for $\mu_0=H_T/2$ and the NLO NNPDF3.1 PDF set. The theoretical
        uncertainties coming from the 7-point scale variation are also shown.}
    \label{tab:ttwNjet2}
\end{table}
\endgroup
% =============================================

As a bonus of our study, in Table \ref{tab:ttwNjet} and  Table \ref{tab:ttwjNjet}  we give results for integrated fiducial cross sections at NLO in QCD for both processes afresh but this time as a function of the number of resolved light jets, denoted as $N_j$.  Also presented  in Table \ref{tab:ttwNjet2} are the results for the $pp \to \WWW +X$ process as obtained with $\mu_0=H_T/2$. In addition, we provide the theoretical uncertainties coming from the 7-point scale variation.  A few comments are in order. First, $N_j$ corresponds to the number of resolved light jets that pass all the cuts  listed in Section \ref{sec:inputparameters}. Thus, in the case of $pp \to \WWW$ ($pp \to \WWW\, j$) the result with $N_j=1$ ($N_j=2$) is described with LO precision only. However, the value given there is obtained with the  NLO input parameters (i.e. NLO PDFs, two-loop running of $\alpha_s$ and for $\Gamma_t^{\rm NLO}$). Second,  the exclusive $pp \to \WWW$ cross section with $N_j=0$ has no real predictable power due to the very bad convergence. Indeed, theoretical uncertainties for  $\sigma^{t\bar{t}W^+}_{H_T/3}(N_j=0)$ as well as for  $\sigma^{t\bar{t}W^+}_{H_T/2}(N_j=0)$ calculated with $\mu_0=H_T/3$ and $\mu_0=H_T/2$  are $84\%$ and  $48\%$ respectively, which should be compered to  $24\%$ LO uncertainties for the $pp \to \WWW$ process. This tells us that the $p_T^{veto}$ cut of $25$ GeV is too low. Third, there are big cancellations between the NLO QCD exclusive $N_j=0$ and $N_j=1$ samples for various values of $\mu_R=\xi\mu_0$ and $\mu_F=\xi\mu_0$ and only the final inclusive $\sigma^{t\bar{t}W^+}_{H_T/3} (N_j\ge 0)$ result has the anticipated reduced scale dependence.  Similarly, for the $pp \to \WWW\, j+X$ process, we can observe big cancellations between the NLO QCD exclusive $N_j = 1$ and $N_j = 2$ results, while the inclusive  $\sigma^{t\bar{t}W^+j}_{H_T/2} (N_j\ge 1)$ prediction has the required reduced NLO scale dependence of $8\%$. Finally, even  for  $\sigma^{t\bar{t}W^+ j}_{H_T/2}(N_j=1)$,  i.e. in the case where the second light jet if resolved and having a $p_T$ above $25$ GeV is vetoed, substantial scale uncertainties ($48\%$) are visible. The latter are larger than LO theoretical uncertainties ($38\%$) obtained for the $pp \to \WWW\, j$ process. While the inclusive NLO results for $pp \to \WWW+X$ and $pp \to \WWW\, j+X$ have the expected reduced theoretical uncertainties, the observed large scale dependence for exclusive NLO samples gives rise to concern as jet vetoes are widely used in experimental analyses at the LHC. For example vetoing  jet activity is crucial in suppressing various background processes or enabling new physics searches. In many LHC analyses the most common jet veto scheme consists of  imposing a maximum transverse momentum cut, $p_T^{veto}$, on anti-$k_T$ jets. For both processes that we are analysing and with our rather inclusive set of cuts the introduction of a jet veto  of $p_T^{veto} = 25$ GeV  introduces large logarithms. In order to reduce theoretical uncertainties, the latter would need to be resummed using approaches similar to those outlined in  Refs. \cite{Berger:2010xi,Stewart:2010tn,Becher:2012qa,Banfi:2012yh,Banfi:2012jm,Michel:2018hui,Campbell:2023cha}.   

From the experimental point of view  $pp \to \WWW$ and $pp \to \WWW\, j$  comprise  similar
final states: three charged leptons, two bottom flavoured jets and the missing transverse momentum from undetected neutrinos, which might be further accompanied by light jets. We would like to compare these two processes also at the differential cross section level in order to see the impact of the additional jet activity on the kinematics of the $\WWW$ final state. Accordingly, in the next step we present differential cross section distributions at NLO in QCD as a function of a few observables that are constructed from three charged leptons and two $b$-jets. In each case upper panels show normalised predictions together with their corresponding uncertainty bands. Bottom panels display the ratio to the prediction for $pp\to \WWW +X$ together with the corresponding uncertainties.  All results are provided for the $H_T$ based scale setting and are obtained with the NLO NNPDF3.1 PDF set.
%
% =============================================
  \begin{figure}[t]
  \begin{center}
    \includegraphics[width=0.49\textwidth]{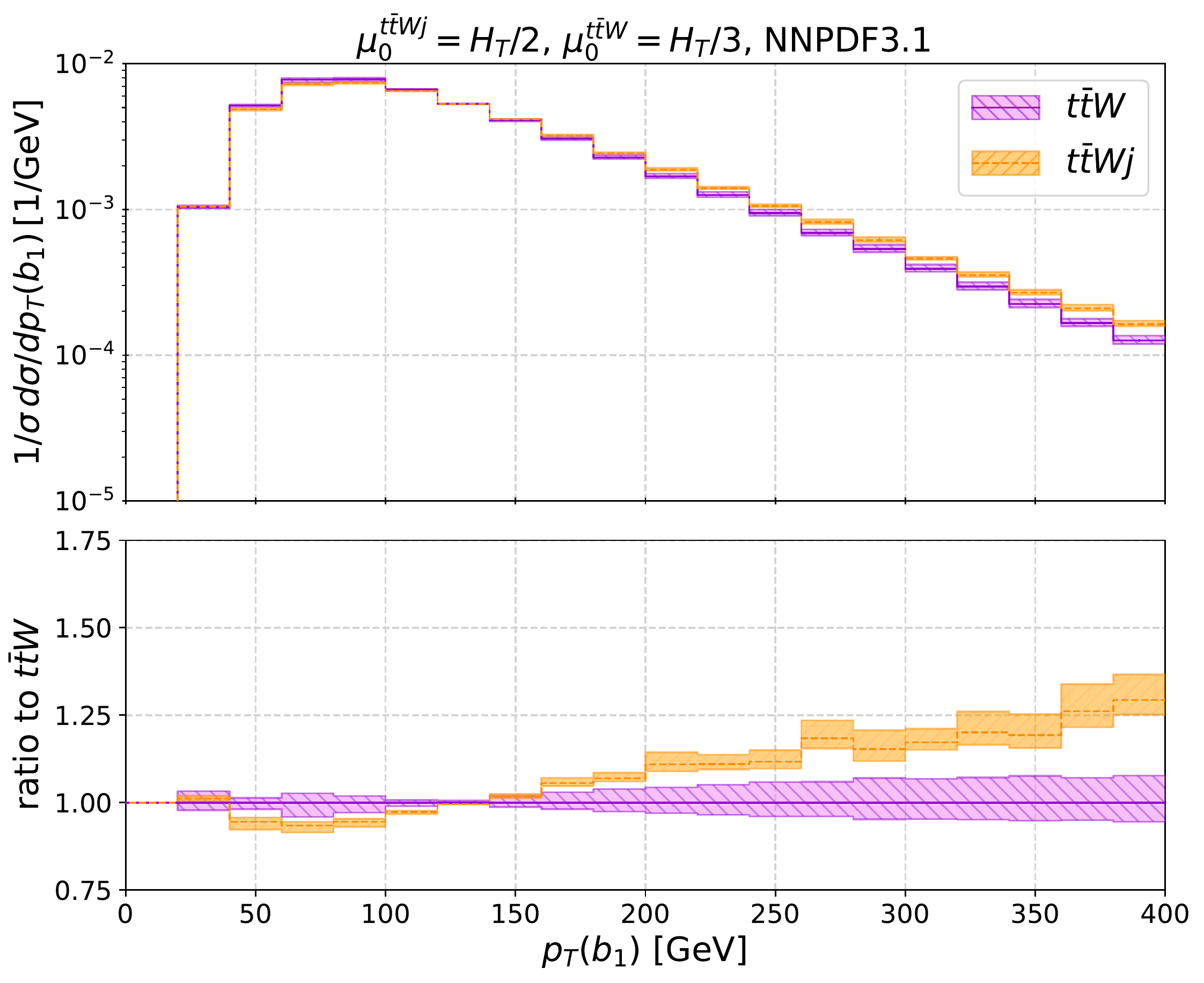}
    \includegraphics[width=0.49\textwidth]{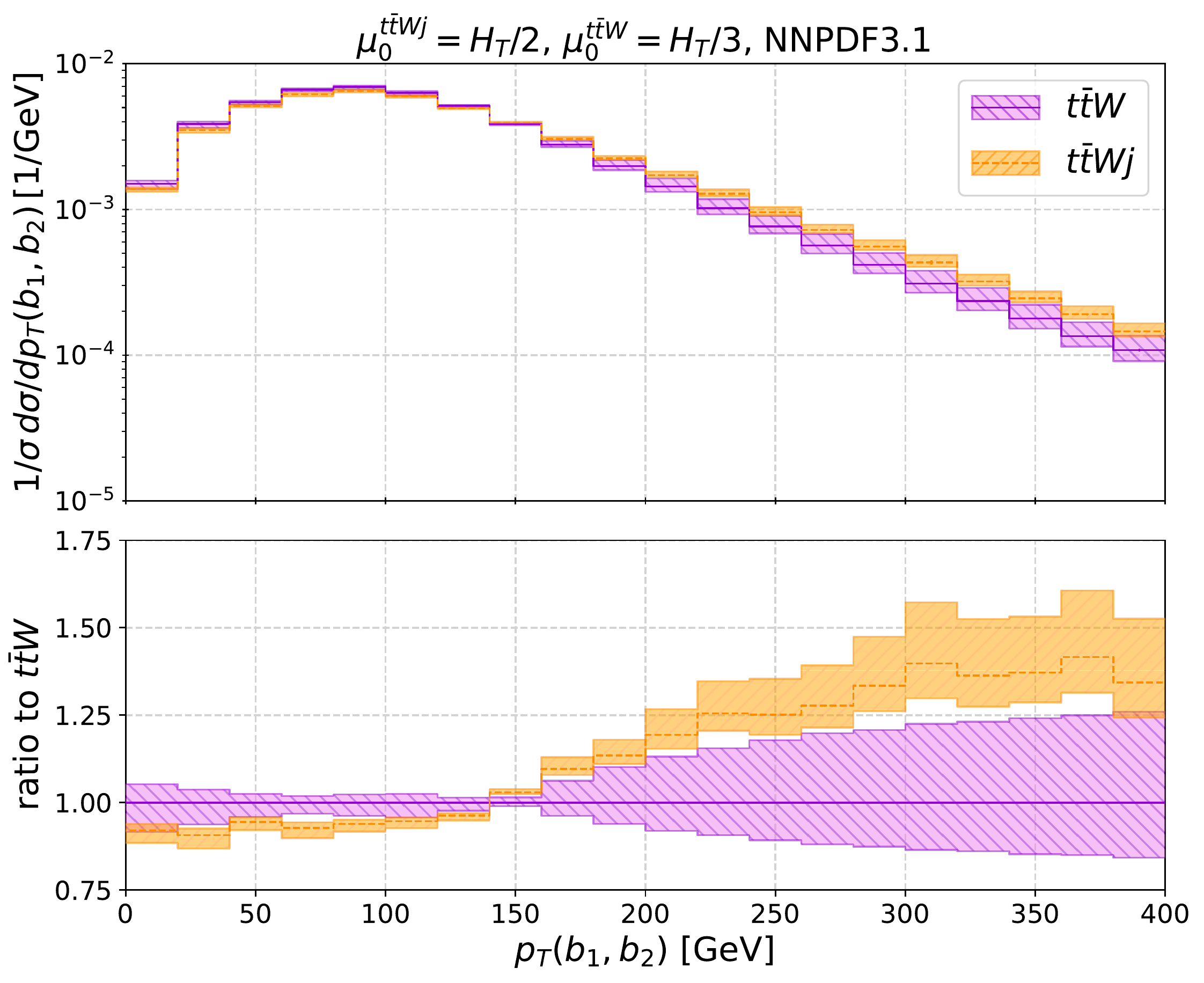}
\end{center}
\caption{\label{fig:ttWcomp_ptb1_ptbb} \it Differential cross section distributions at NLO in QCD as a function of  $p_T(b_1)$ and $p_T(b_1,b_2)$  for the $pp \to \WWW \, +X$ and $pp \to \WWW \, j+X$ process 
at the LHC with $\sqrt{s}= 13$ TeV.  Upper panels show normalised 
predictions together with their corresponding uncertainty bands.
  Bottom panels display the ratio to the $pp \to \WWW \, +X$ prediction.  All results are provided for $\mu_0=H_T/2$  in the case of the $pp \to \WWW \, j+X$ process 
 and with $\mu_0=H_T/3$ for $pp \to \WWW \, +X$. Theoretical predictions are obtained for the NLO NNPDF3.1 PDF set. }
 \end{figure}
  % =============================================
  \begin{figure}[t]
  \begin{center}
    \includegraphics[width=0.49\textwidth]{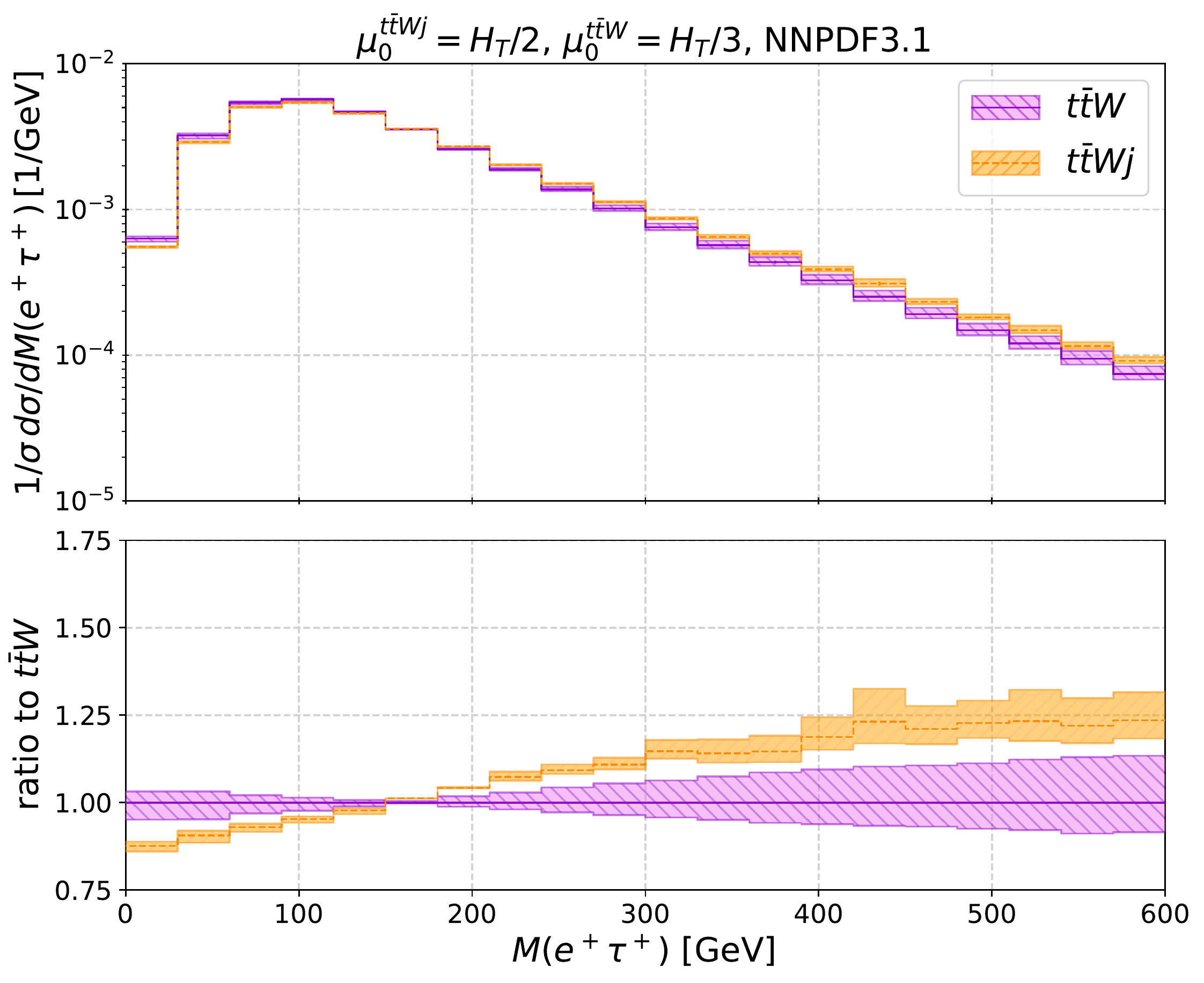}
    \includegraphics[width=0.49\textwidth]{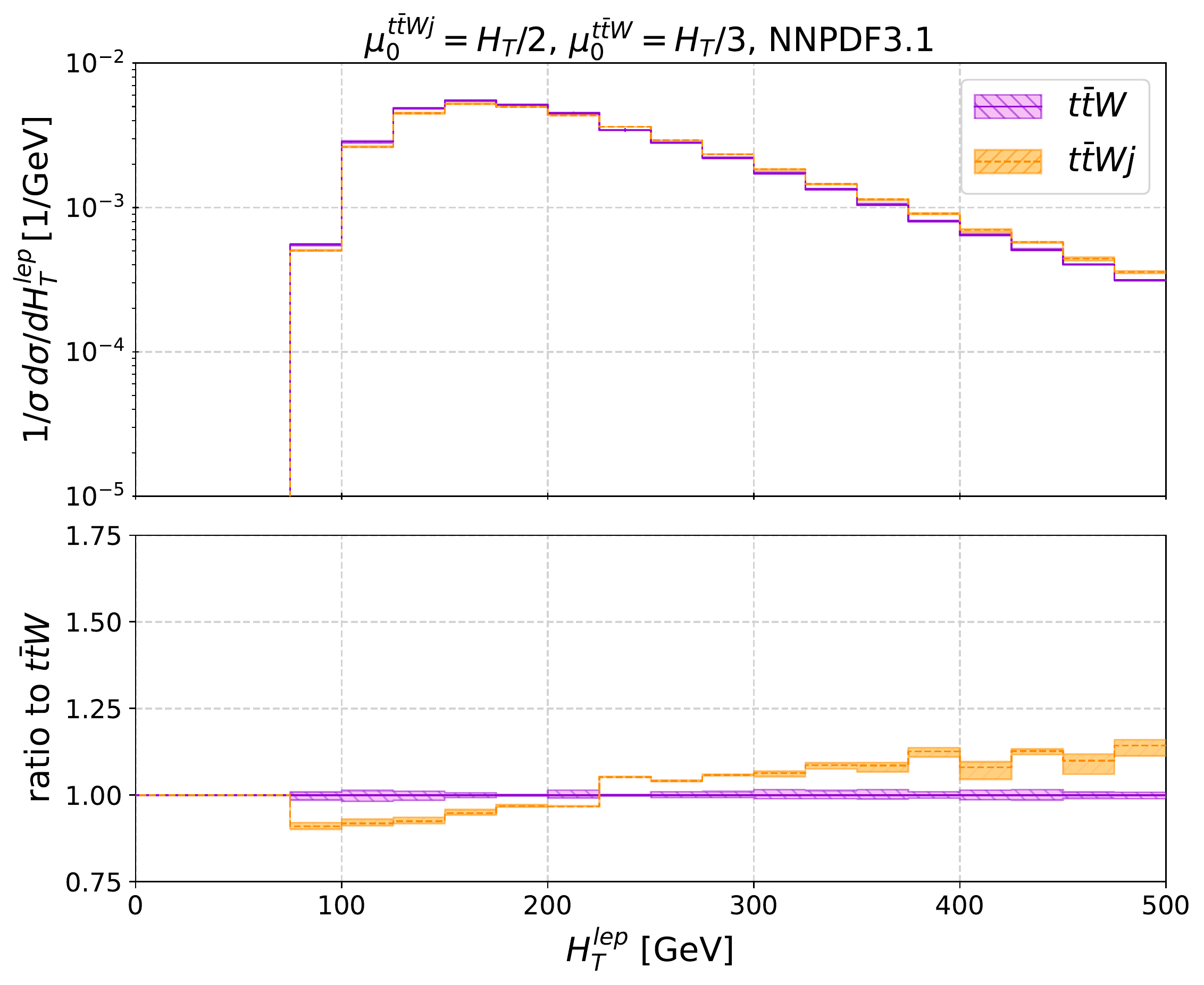}
\end{center}
\caption{\label{fig:ttWcomp_melta_htlep} \it   Differential cross section distributions at NLO in QCD as a function of  $M(e^+\tau^+)$ and $H_T^{lep.}$  for the $pp \to \WWW \, +X$ and $pp \to \WWW \, j+X$ process 
at the LHC with $\sqrt{s}= 13$ TeV.  Upper panels show normalised 
predictions together with their corresponding uncertainty bands.
  Bottom panels display the ratio to the $pp \to \WWW \, +X$ prediction.  All results are provided for $\mu_0=H_T/2$  in the case of the $pp \to \WWW \, j+X$ process  and with $\mu_0=H_T/3$ for $pp \to \WWW \, +X$. Theoretical predictions are obtained for the NLO NNPDF3.1 PDF set. }
 \end{figure}
 % =============================================
  \begin{figure}[t]
  \begin{center}
    \includegraphics[width=0.49\textwidth]{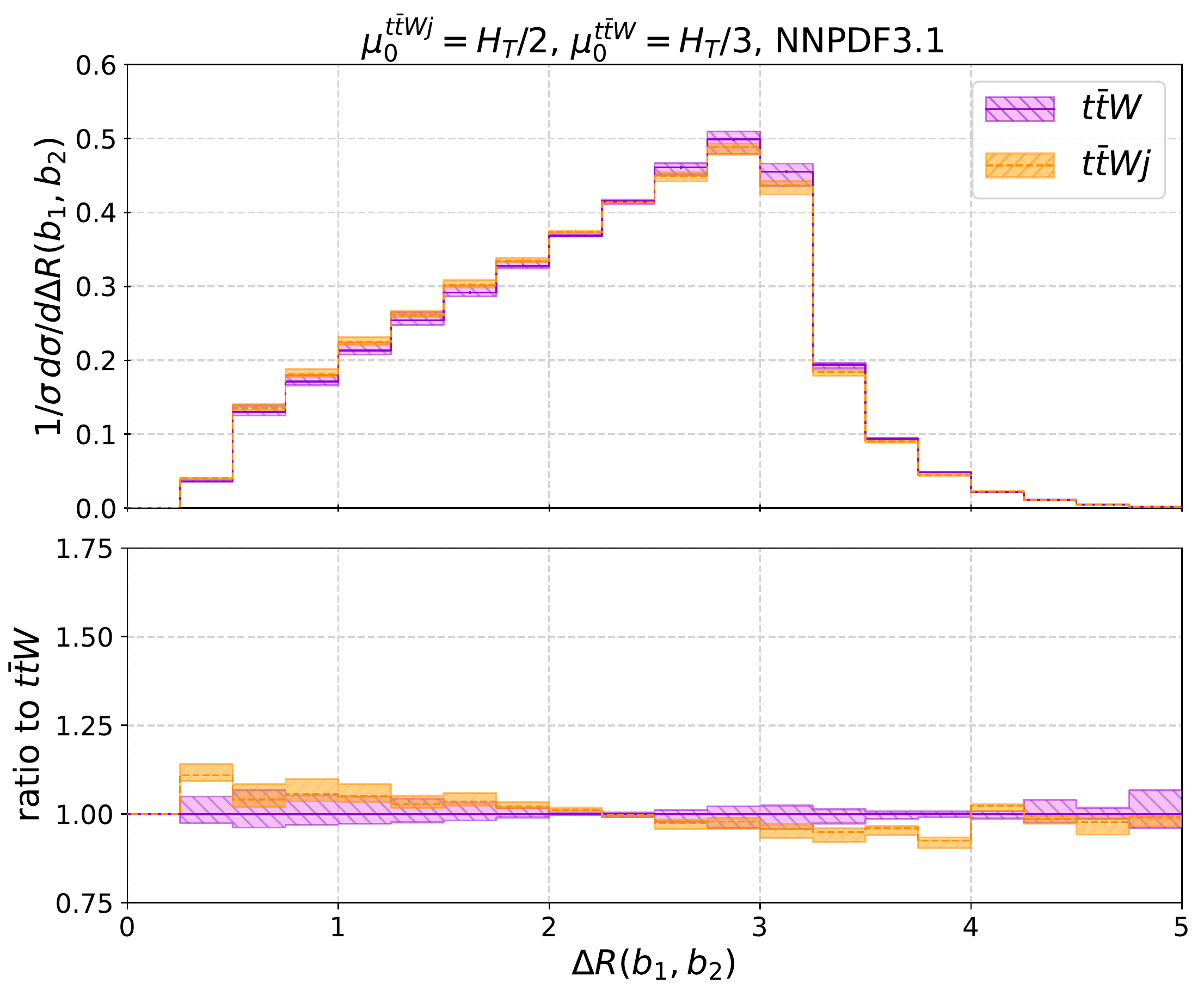}
    \includegraphics[width=0.49\textwidth]{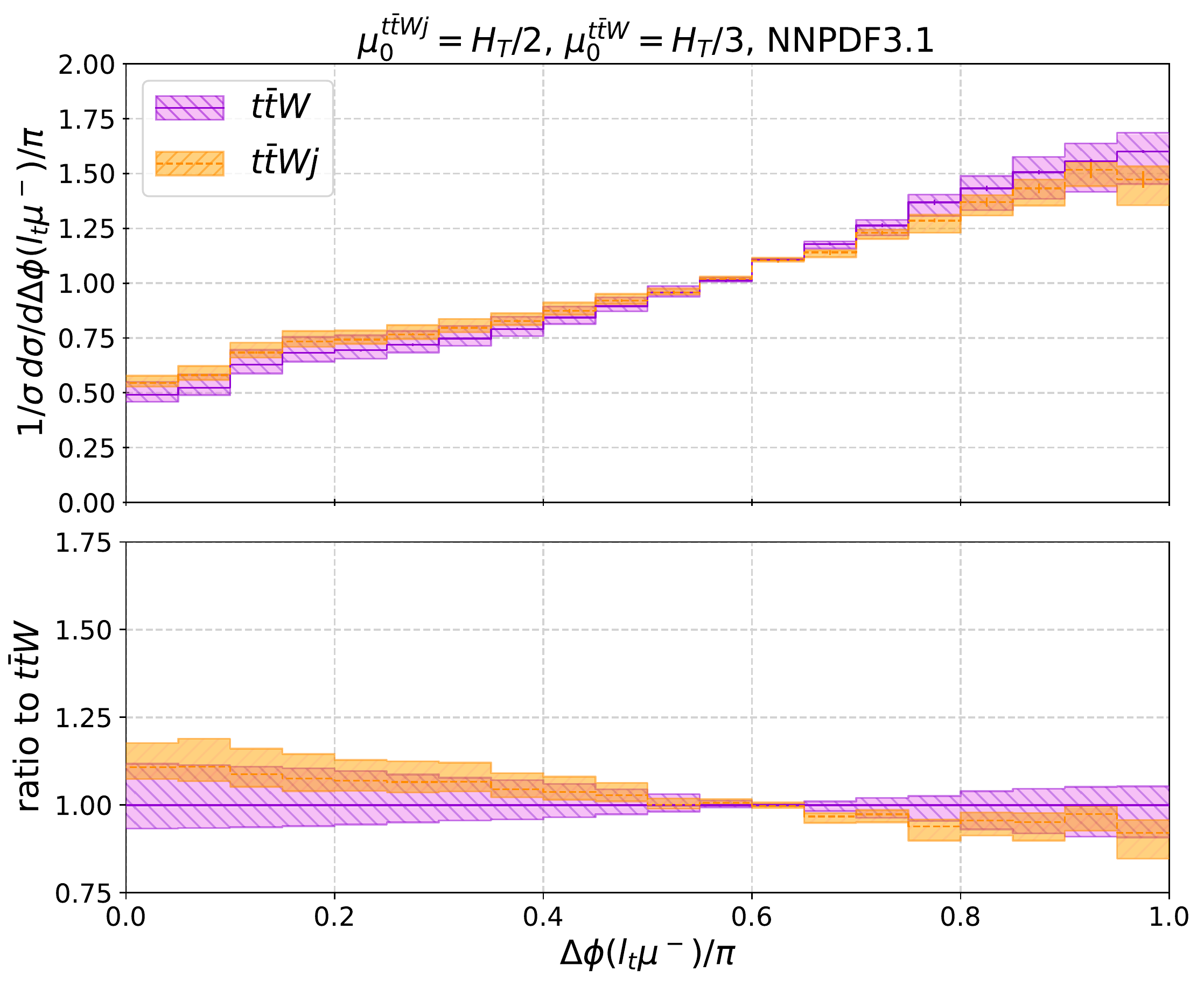}
\end{center}
\caption{\label{fig:ttWcomp_drbb_dphiltmu} \it
  Differential cross section distributions at NLO in QCD as a function of  $\Delta R(b_1,b_2)$ and $\Delta\phi(l_t\mu^-)$  for the $pp \to \WWW \, +X$ and $pp \to \WWW \, j+X$ process 
at the LHC with $\sqrt{s}= 13$ TeV.  Upper panels show normalised 
predictions together with their corresponding uncertainty bands.
  Bottom panels display the ratio to the $pp \to \WWW \, +X$ prediction.  All results are provided for $\mu_0=H_T/2$  in the case of the $pp \to \WWW \, j+X$ process  and with $\mu_0=H_T/3$ for $pp \to \WWW \, +X$. Theoretical predictions are obtained for the NLO NNPDF3.1 PDF set. }
 \end{figure}
 % =============================================

In the first step we present observables related to the $b$-jet kinematics.  In Figure \ref{fig:ttWcomp_ptb1_ptbb} we display the transverse momentum of the hardest $b$-jet and  the $b\bar{b}$ system denoted as $p_T(b_1)$ and  $p_T(b_1,b_2)$ respectively. For $p_T(b_1)$ we can see substantial differences between the two processes up to $30\%$, which are well above the theoretical  uncertainties due to the scale dependence. The latter are very similar for both processes and are maximally up to $6\%-8\%$. In the case of $p_T(b_1,b_2)$ substantial theoretical uncertainties up to even $25\%$ are obtained for $pp \to \WWW +X$, while in the case of $pp \to \WWW\, j+X$ they are reduced to $14\%$. However, the difference between the two processes is up to $42\%$, so it is still clearly outside the uncertainty bands.  

This general behaviour is also present for observables constructed from  the
charged leptons. In Figure \ref{fig:ttWcomp_melta_htlep} we show the invariant mass of the two same-sign leptons, $M(e^+\tau^+)$, together with  the scalar sum of the transverse momenta of the three charged leptons denoted as $H_T^{lep}$ and given by 
\begin{equation}
H_T^{lep} = p_T(e^+) + p_T(\mu^-) +p_T(\tau^+) \,.
\end{equation}
For the $M(e^+\tau^+)$ observable, which is very relevant in various beyond SM physics scenarios, the difference between the two processes is up to $24\%$. At the same time, the shape distortion  for this observable is much larger of the order of $36\%$. Theoretical uncertainties are in the range of $8\%-13\%$ in the tail of the distribution and as small as $2\%-5\%$ at the beginning of the $M(e^+\tau^+)$ spectrum.  For our final dimensionful observable, $H_T^{lep}$, a similar pattern can be observed, albeit the shape distortion is diminished and only of the order of $23\%$. Meanwhile theoretical uncertainties are smaller in the range of $2\%-5\%$.  

Lastly we want to analyse two dimensionless quantities. In Figure
\ref{fig:ttWcomp_drbb_dphiltmu} we present observables that are related to
the underlying basic process they have in common, i.e. top-quark pair production. Specifically, we show normalized distributions connected to the top-quark kinematics, namely the $\Delta R(b_1,b_2)$ separation  between  the two $b$-jets and the angular difference $\Delta \phi (\ell_t \mu^-)$  between the two leptons coming from top-quark decays.  For  $\Delta \phi (\ell_t \mu^-)$ we use
the prescription outlined in Section \ref{sec:inputparameters}  to  identify the positively charged lepton originating from $t \to bW^+  \to b \ell^+ \nu_\ell$, referred to  as $\ell_t$. The second one is uniquely identified via  $\bar{t} \to \bar{b} W^- \to \bar{b} \mu^- \bar{\nu}_\mu$. Also for these two observables we observe large shape distortions. In detail, in both cases they are of the order of $20\%$. This should be contrasted with theoretical errors which are in the range of $4\%-7\%$ for $\Delta R(b_1,b_2)$ and up to $8\%-12\%$ for $\Delta \phi (\ell_t \mu^-)$. Due
to the additional recoil of the extra light jet more collinear configurations  are enhanced while  anti-collinear  are being reduced. The same can be read out from the $\Delta \phi (\ell_t \mu^-)$ differential cross section distribution where small separations between the charged leptons from top-quark decays are increased while events with $\Delta \phi (\ell_t \mu^-) \sim \pi$ are decreasing for the $pp \to \WWW\, j+X$ process. 

% =============================================
%
\section{Summary}
\label{summary}
%
% =============================================
%

In this paper we calculated NLO QCD corrections to the $pp \to \WWW\, j+X$ process at the LHC with $\sqrt{s} = 13$ TeV. In the computation off-shell top quarks and massive gauge bosons have been described by Breit-Wigner propagators, furthermore double-, single- and non-resonant top-quark contributions have been consistently incorporated already at the matrix element level. We have evaluated the integrated fiducial cross section and its scale dependence  for a few scale choices, i.e. for the fixed 
scale  setting $\mu_R=\mu_F=\mu_0 = m_t + m_W /2$ and for two  dynamical 
scale choices $\mu_0=H_T/2$ and $\mu_0=E_T/2$. The effect of the NLO QCD corrections on the integrated cross section is small to moderate depending on the choice of PDFs. For our default setup with $\mu_0=H_T/2$ and the NNPDF3.1 PDF set the full $pp$ cross section receives moderate and positive NLO corrections of $23\%$. The theoretical uncertainties resulting from scale variations are $8\%$ at NLO. The internal NNPDF3.1 PDF uncertainties are of the order of $1\%$. They are, therefore, well below the theoretical uncertainties due to the 7-point scale variation.  The latter remains the dominant source of the theoretical systematics.

The impact of NLO QCD corrections on differential cross section 
distributions has been additionally analysed.  For dimensionful observables large NLO QCD corrections have been obtained up to even $80\% - 90\%$ in specific phase-space regions.  This shows the importance of higher-order corrections in describing additional light jet activity in the $pp \to \WWW\, j+X$ process. The corresponding  theoretical uncertainties are maximally up to $18\%$. This value should be compared to the LO uncertainties which are  at a rather constant  level of $40\%-45\%$ for this process. Furthermore, we have verified that the PDF uncertainties are negligible at the differential cross section level when comparing to the theoretical uncertainties due to the scale dependence.

Finally, we have investigated the importance of additional jet activity in the $pp \to t\bar{t}W^\pm$ process. We have compared both processes $pp \to \WWW\, j+X$ with $pp \to \WWW +X$. For $p_T(j_1) > 25$ GeV and our inclusive cuts, more than half of $pp \to \WWW +X$ events are expected to be accompanied by an additional hard light jet. Specifically, the size of the $pp \to \WWW\, j+X$ contribution in the more inclusive $pp \to \WWW +X$ cross section has been estimated to be at the level of $50\% - 70\%$ depending on the scale choice and the order in the perturbative expansion in $\alpha_s$. At the differential cross section level significant differences between the two processes could be seen as well as large shape distortions due to the presence of the extra light jet. The correct description of the $pp \to t\bar{t}W^\pm j$ process is therefore essential in the study of $pp \to t\bar{t}W^\pm$ production to achieve more precise theoretical predictions. 

\acknowledgments{
The work of  M.R. and M.W.  was supported by the Deutsche
Forschungsgemeinschaft (DFG) under grant 396021762 $-$
TRR 257: {\it P3H - Particle Physics Phenomenology after the Higgs
Discovery}. Support by a grant of the Bundesministerium f\"ur Bildung
und Forschung (BMBF) is additionally acknowledged.

The research of H.Y.B. was supported by the China Postdoctoral Science Foundation under Grant No. 2022TQ0012
  and No. 2023M730097.

Simulations were performed with computing resources granted by RWTH
Aachen University under projects {\tt p0020216} and {\tt rwth0414}. }

%\bibliography{references} 

%\bibliographystyle{JHEP}

\providecommand{\href}[2]{#2}\begingroup\raggedright\endgroup

\end{document}